\documentclass[a4paper,amsmath,amssymb,prb,preprintnumbers,superscriptaddress,twocolumn,showkeys,longbibliography]{revtex4-2}

\pdfoutput=1 

\usepackage{graphicx}

\usepackage{dcolumn}  
\usepackage{bm}         
\usepackage{amssymb}
\usepackage{amsmath}
\usepackage{epsfig}
\usepackage{xcolor}
\usepackage{soul} 

\definecolor{pinegreen}{rgb}{0.0, 0.47, 0.44}
\definecolor{persiangreen}{rgb}{0.0, 0.65, 0.58}
\definecolor{pakistangreen}{rgb}{0.0, 0.4, 0.0}
\definecolor{mossgreen}{rgb}{0.68, 0.87, 0.68}
\definecolor{olivegreen}{rgb}{0.58, 0.57, 0.27}
\definecolor{yelloworange}{rgb}{1.0, 0.68, 0.26}
\definecolor{purple}{rgb}{0.63, 0.13, 0.94}
\definecolor{brickred}{rgb}{0.8, 0.25, 0.33}
\definecolor{palecarmine}{rgb}{0.69, 0.25, 0.21}
\definecolor{bondiblue}{rgb}{0.0, 0.58, 0.71}
\definecolor{awesome}{rgb}{1.0, 0.13, 0.32}
\definecolor{carnationpink}{rgb}{1.0, 0.65, 0.79}
\definecolor{airforceblue}{rgb}{0.36, 0.54, 0.66}
\definecolor{beaublue}{rgb}{0.74, 0.83, 0.9}
\definecolor{bleudefrance}{rgb}{0.19, 0.55, 0.91}
\definecolor{blue(pigment)}{rgb}{0.2, 0.2, 06}
\definecolor{brightlavender}{rgb}{0.75, 0.58, 0.89}
\definecolor{brightgreen}{rgb}{0.4, 1.0, 0.0}
\definecolor{caribbeangreen}{rgb}{0.0, 0.8, 0.6}
\usepackage{xspace}

\newif\ifcom
\newif\ifdel
\newif\ifdelsecondv

\comtrue                    %
\deltrue                      %

\delsecondvtrue                                       %

\begin{document}

\title{Optical imaging of strain-mediated phase coexistence during electrothermal switching in a Mott insulator}

\author{Matthias Lange}
\affiliation{%
Physikalisches Institut, Center for Quantum Science (CQ) and LISA$^+$,
Eberhard Karls Universit\"at T\"ubingen,
Auf der Morgenstelle 14,
72076 T\"ubingen, Germany}

\author{Stefan Gu\'enon}
\email{stefan.guenon@uni-tuebingen.de}
\affiliation{%
Physikalisches Institut, Center for Quantum Science (CQ) and LISA$^+$,
Eberhard Karls Universit\"at T\"ubingen,
Auf der Morgenstelle 14,
72076 T\"ubingen, Germany}

\author{Yoav Kalchheim}
\affiliation{%
Center for Advanced Nanoscience, Department of Physics, 
University of California –- San Diego, 
9500 Gilman Drive, La Jolla, CA92093-0319, USA}
\affiliation{%
Department of Materials Science and Engineering, Technion -- Israel Institute of Technology, Technion City, 32000 Haifa, Israel}

\author{Theodor Luibrand}
\affiliation{%
Physikalisches Institut, Center for Quantum Science (CQ) and LISA$^+$,
Eberhard Karls Universit\"at T\"ubingen,
Auf der Morgenstelle 14,
72076 T\"ubingen, Germany}

\author{Nicolas Manuel Vargas}
\affiliation{%
Center for Advanced Nanoscience, Department of Physics, 
University of California –- San Diego, 
9500 Gilman Drive, La Jolla, CA92093-0319, USA}

\author{Dennis Schwebius}
\affiliation{%
Physikalisches Institut, Center for Quantum Science (CQ) and LISA$^+$,
Eberhard Karls Universit\"at T\"ubingen,
Auf der Morgenstelle 14,
72076 T\"ubingen, Germany}

\author{Reinhold Kleiner}
\affiliation{%
Physikalisches Institut, Center for Quantum Science (CQ) and LISA$^+$,
Eberhard Karls Universit\"at T\"ubingen,
Auf der Morgenstelle 14,
72076 T\"ubingen, Germany}

\author{Ivan K.~Schuller}
\affiliation{%
Center for Advanced Nanoscience, Department of Physics, 
University of California –- San Diego, 
9500 Gilman Drive, La Jolla, CA92093-0319, USA}

\author{Dieter Koelle}
\affiliation{%
Physikalisches Institut, Center for Quantum Science (CQ) and LISA$^+$,
Eberhard Karls Universit\"at T\"ubingen,
Auf der Morgenstelle 14,
72076 T\"ubingen, Germany}


\begin{abstract} 

Resistive-switching -- the current-/voltage-induced electrical resistance change -- is at the core of memristive devices, which play an essential role in the emerging field of neuromorphic computing.
This study is about resistive switching in a Mott-insulator, which undergoes a thermally driven metal-to-insulator transition. 
Two distinct switching mechanisms were reported for such a system: electric-field-driven resistive switching and electrothermal resistive switching.
 The latter results from an instability caused by Joule heating. 
Here, we present the visualization of the reversible resistive switching in a planar V$_2$O$_3$ thin-film device using high-resolution wide-field microscopy in combination with electric transport measurements. 
We investigate the interaction of the electrothermal instability with the strain-induced spontaneous phase-separation in the V$_2$O$_3$ thin film at the Mott-transition. 
The photomicrographs show the formation of a narrow metallic filament with a minimum width  $\lesssim$ 500\,nm.
Although the filament formation and the overall shape of the current-voltage characteristics (IVCs) are typical of an electrothermal breakdown, we also observe atypical effects like oblique filaments, filament splitting, and hysteretic IVCs with sawtooth-like jumps at high currents in the low-resistance regime. 
We were able to reproduce the experimental results in a numerical model based on a two-dimensional resistor network. 
This model demonstrates that resistive switching in this case is indeed electrothermal and that the intrinsic heterogeneity is responsible for the atypical effects. This heterogeneity is strongly influenced by strain, thereby establishing a link between switching dynamics and structural properties.
\end{abstract} 


\maketitle

The strongly correlated electron system V$_2$O$_3$ is a prototypical Mott-Hubbard insulator \cite{Limelette03}. 
At room temperature, stoichiometric V$_2$O$_3$ is a paramagnetic metal with corundum structure, which undergoes a metal-to-insulator-transition (MIT) in cooling below about 160\,K.
The insulating phase is antiferromagnetic with monoclinic structure.
Upon heating, the insulating phase undergoes a thermally driven insulator-to-metal-transition (IMT) \cite{McWhan70,Imada98,Singer18}.

In recent years, there has been a growing interest in utilizing the MIT/IMT in devices based on strongly correlated oxides \cite{Yang11}, e.g., as memristors \cite{delValle17,Janod15}, electro-optical elements \cite{Markov15,Butakov18} and building blocks for neuromorphic computing \cite{Pickett13,Schuller15,Stoliar17,delValle19}. 
The basis for these applications is resistive switching, in which an applied electric current or field induces the IMT \cite{Lin17}.

Several mechanisms were suggested for resistive switching in Mott insulators:
At very high fields ($\gtrsim 100\,$MV/m) Landau-Zener tunneling across the Mott gap increases the free carrier concentration which eventually destabilizes the insulating state, leading to the IMT \cite{Oka03,Yamakawa17,Oka05,Eckstein10,Hedrich-Meisner10}.
However, switching with considerably lower fields has been observed in some cases \cite{Kumai99,Yamanouchi99,Valmianski18}.
The low-field switching was attributed to either mid-gap tunneling \cite{Lee14,Sugimoto08}, an electric-field-driven Mott-gap collapse \cite{Mazza16} or a spatially inhomogeneous metal-insulator mixed state \cite{Li15}.

In addition to these mechanisms, in materials exhibiting a thermally driven IMT, a universal electrothermal breakdown must be considered.
This instability is not directly related to the Mott-Hubbard physics, but is the result of an electrothermal instability created by the strong temperature dependence of the electric resistivity at the IMT \cite{Duchene71,Guenon13,Brockman14,Polozov20}.
Such an electrothermal instability can lead to resistive switching:
When current passes through a highly resistive (insulating) device, Joule heating increases the device temperature.
Consequently, small spatial variations in the current density may lead to spatial variations of the local device temperature.
The sharp decrease in the resistivity with increasing temperature (at the IMT) acts as a positive feedback mechanism, which amplifies the evolution of a spatially inhomogeneous state.
For electric currents above a certain threshold, this may lead to a runaway effect where more and more current is concentrated in a small section of the device, considerably increasing the local temperature.
This runaway effect results in the formation of a highly conductive (metallic) filament connecting the device electrodes.
The filament is sustained in the metallic state by the Joule heating concentrated within it, while its surroundings remain in the insulating state at the lower base temperature.
In this case, resistive switching is the result of current and temperature redistribution in the device (see Ref.~\onlinecite{Gurevich87} for an extensive discussion of this phenomenon).

Considering V$_2$O$_3$ devices, it is possible to suppress electrothermal effects and enter a purely electronic regime in ultrafast pump-probe experiments \cite{Giorgianni19}.
In a study on V$_2$O$_3$ nanodevices, where size effects and electrode cooling reduced Joule heating, evidence for a dielectric breakdown was found \cite{Valmianski18}, and in a recent study, the authors found that defects in the V$_2$O$_3$ thin films enhance the efficiency of field-assisted carrier generation and considerably reduce the threshold for a dielectric (non-thermal) breakdown \cite{Kalcheim20}. 
Electrothermal switching is therefore expected in large and highly pure devices where cooling from the contacts is inefficient and defects do not play a substantial role in the switching process.
Moreover, we note that due to the elastic strain and coupling between the structural and the electronic degrees of freedom, V$_2$O$_3$ thin films on single crystal substrates have very rich physics on their own \cite{Schuler97,Alyabyeva18,Ronchi19}. 
Of particular importance is the spontaneous phase separation with herringbone domain structures at the sub-micrometer scale due to strain minimization at the MIT/IMT \cite{McLeod17}; this in turn is likely to also affect resistive switching at the IMT.

Despite major research efforts, details of the resistive switching mechanism remain elusive. 
In particular, for systems with a thermally driven IMT, the question of purely electrically vs.~electrothermally induced resistive switching has been debated over recent years \cite{Brockman14, Guenon13, Valmianski18, Janod15, Stoliar13, Diener18, Gopalakrishnan09, Li15, Kalcheim20, Ronchi21}.
A key problem for understanding voltage- or current-induced switching is based on the fact that in experiments the switching is analyzed by electric transport measurements, while in the case of filament formation, the system is inherently strongly inhomogeneous, eventually including a strongly inhomogeneous spatial distribution of current density and temperature.
Because of this, it should be highly rewarding to combine electric transport measurements with imaging of filament formation on the submicrometer scale.
This may, in particular, provide so far inaccessible insights into the relation between strain-induced domain formation and switching-induced filament formation at the IMT/MIT.

In this study, we combine electric transport measurements with cryogenic optical wide-field microscopy (with a spatial resolution of about $0.5\,\mu$m) on V$_2$O$_3$ thin film devices. 
Due to different reflectivity, the photomicrographs yield a strong contrast between metallic and insulating regions, which allows for imaging of the metal-insulator spontaneous phase separation at the MIT/IMT and for the determination of the spatial distribution of local MIT and IMT temperatures. 
Furthermore, we image the evolution of the formation of highly conducting (metallic) filaments that accompany resistive switching upon sweeping an applied current in the insulating state and recording simultaneously the current-voltage characteristics (IVCs). 
This way, characteristics of the electric transport measurements can be related to the formation, growth and extinction of metallic filament features at the submicron scale. 
We developed a numerical model based on a two-dimensional resistor network, which takes into account the experimentally determined spatial distribution of the local MIT/IMT temperatures. 
With this model, we calculate IVCs and the spatial distribution of current density and temperature upon electrothermal resistive switching and find excellent agreement with experimentally determined IVCs and optically detected filament formation. 
These results provide a profound understanding of the resistive-switching-mechanism details at play.
In particular, comparing the spontaneously formed metal-insulator domain structure with the filamentary structures resulting from electrothermal switching also reveals a hitherto unknown interplay between strain and switching.

\section*{Results and Discussion}
\label{sec:Results}

This section is organized as follows: 
First, we discuss the thermally driven MIT/IMT in the planar V$_2$O$_3$ thin-film device under investigation. 
In addition to resistance $R$ vs temperature $T$ data, we present photomicrographs, which show characteristic insulator-metal herringbone domain structures at the MIT/IMT. 
We discuss how to derive the local MIT/IMT temperature for individual pixels from a photomicrograph temperature series, which leads to maps of "Spatially Resolved Transition  Temperatures". 
The main results are presented in the section "Current-Induced IMT -- Resistive Switching". 
There, we present IVCs at a base temperature $T_{\rm b}=158\,$K where the device is in the insulating state close to the IMT.
The current-biased device shows switching to a low-resistance state upon ramping up and switch-back to a high-resistance state upon ramping down the current. 
Photomicrographs simultaneously acquired at various bias points on the IVC show that the switching to low- and high-resistance states is accompagnied with the generation and extinction of metallic filaments, respectively.
We argue that the data have many characteristics of an electrothermal breakdown, including some effects that are atypical for a breakdown in a homogeneous medium without thermal hysteresis.  
On the basis of a two-dimensional (2D) resistor network model, which includes the experimentally determined spatial distribution of MIT/IMT transition temperatures, we performed numerical simulations of IVCs and spatial distribution of current density and local temperature.
The comparison of experimental data and numerical simulations allows us to explain the resistive switching details, including the evolution of the filamentary structures with current during resistive switching.

\begin{figure*}[t]
\includegraphics{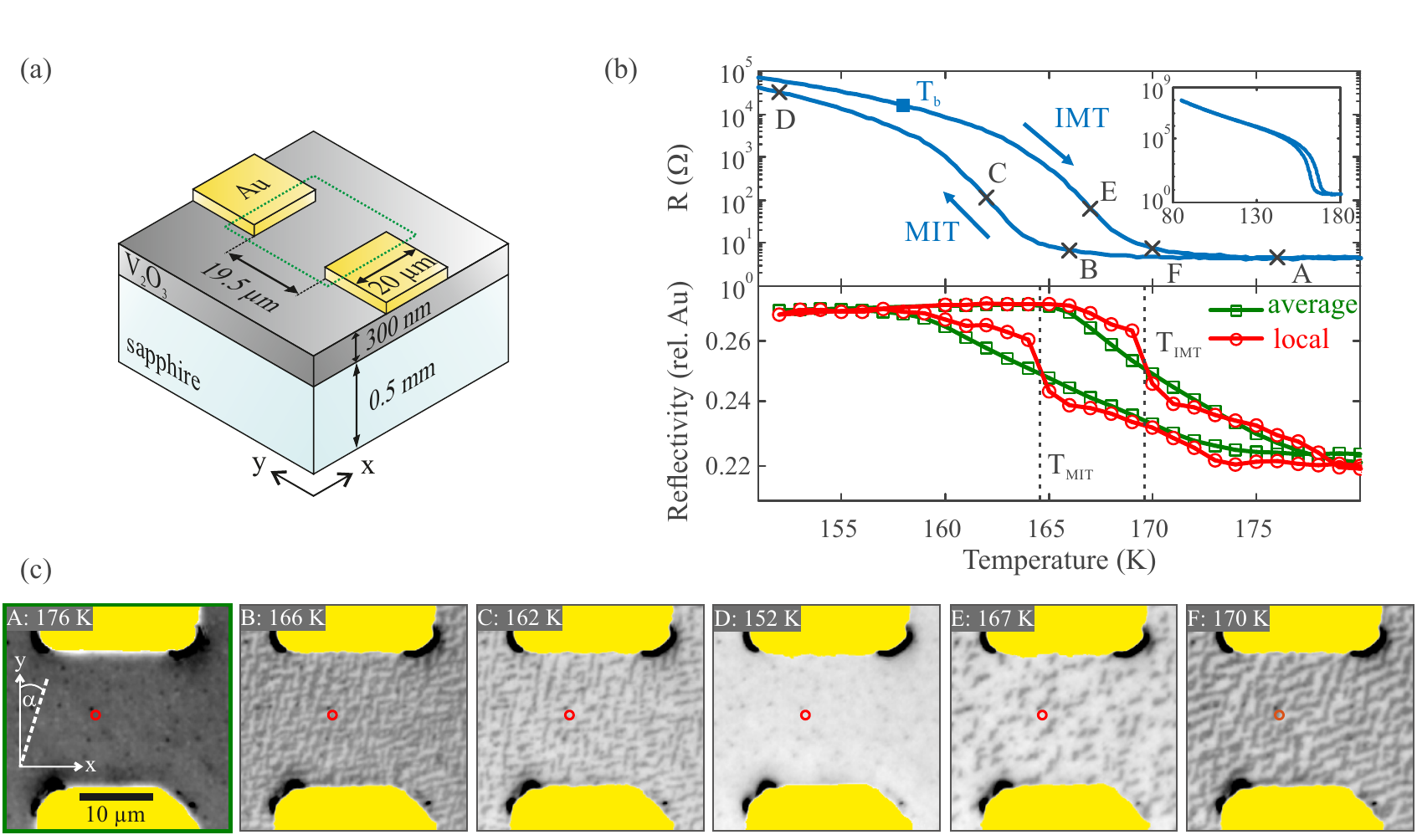}
\caption{
(a) Schematic view of the planar V$_2$O$_3$ device under investigation.
The two Au electrodes serve as combined current/voltage taps for two-probe electric transport measurements.
The dotted green line indicates the area of interest (field of view of the images shown in (c)).
(b) Resistively and optically detected MIT/IMT.
UPPER PANEL: Electrical resistance $R(T)$ (measured with $I=5\,\mu$A);
arrows indicate sweep direction of temperature $T$.
Data points marked by letters correspond to images shown in (c).
$T_{\rm b}$ indicates the base temperature set for the measurements presented in Figure \ref{Fig3}. 
INSET: $R(T$) including the lowest temperature of the thermal cycle.
LOWER PANEL: Reflectivity vs $T$ of the V$_2$O$_3$ film normalized to the reflectivity of the Au electrodes.
The green curve represents the average over the area of interest, while the red curve represents the reflectivity of a single pixel at the location indicated by the red circles in (c).
Vertical dashed lines indicate the local transition temperatures $T_{\rm MIT}$ and $T_{\rm IMT}$, determined for the single pixel.
(c) Selection of photomicrographs acquired during one thermal cycle.
Au electrodes are indicated by yellow areas.}
\label{Fig1}
\end{figure*}

\subsection*{Thermally Driven MIT/IMT.}

\subsubsection*{Device Resistance.}
%
The device under investigation is an unpatterned 300-nm-thick V$_2$O$_3$ thin film,  grown on r-cut sapphire.
The current $I$ is injected via two Au electrodes patterned on top of the film, and the voltage is measured in a two-point configuration (see Figure \ref{Fig1}a and methods section).
Figure \ref{Fig1}b, upper panel, shows the electrical resistance $R$ vs.~temperature $T$ of the device, acquired with a bias current $I=5\,\mu$A.
The large resistance change at the MIT by four orders of magnitude is indicative of a high-quality V$_2$O$_3$-thin-film with few defects.
Due to the first-order nature of the MIT/IMT, the $R$($T$)-curve is hysteretic, with an MIT temperature $T_{\rm MIT}$ about 5\,K below the IMT temperature $T_{\rm IMT}$.
Consequently, for the IVC measurements and imaging presented below, it has been vital to prepare a well-defined initial state in order to obtain reproducible results.
This has been achieved by thermal cycling, i.e., by cooling or by heating to temperatures, where the V$_2$O$_3$ is in a pure insulating or metallic state, before heating or cooling to the targeted temperature.

\subsubsection*{Phase Separation at the MIT/IMT.}

We have acquired a temperature series of wide-field microscopy images during one thermal cycle with a 1\,K stepsize across the MIT/IMT.
Figure \ref{Fig1}c shows a selection of images from this series.
An extended photomicrograph series is included in the supplement (Figure S1).
Capital letters indicate the corresponding temperatures in the $R$($T$) curve (Figure \ref{Fig1}b upper panel).
By averaging over the area of interest, we obtain the temperature dependence of the reflectivity of the V$_2$O$_3$ film, normalized to the reflectivity of the Au electrodes (see Figure \ref{Fig1}b lower panel).
The hysteresis in the $R(T)$ curve is well reproduced by the normalized reflectivity vs $T$.
The contrast in the images is due to the fact that the insulating phase has a significantly higher reflectivity than the metallic phase (see methods).

The photomicrographs acquired in the hysteretic temperature regime show spontaneous insulator/metal phase separation with herringbone-domains \cite{McLeod17}.
Domain patterns, bearing a strong resemblance to those observed here, can be found by numerical approximations to the Cahn-Larch\'e equation, which describes the phase separation of a binary mixture in the presence of elastic stress (see Figure 6 in ~\cite{Garcke05}).
A fast Fourier transform (FFT) analysis of the domain pattern is included in the supplement (Figure S2).
The FFT analyisis reveals that the domains have two preferred directions; one is at an angle $\alpha\approx 9\,^\circ$ at the MIT, which switches to $\alpha\approx 22\,^\circ$ at the IMT, and the second is at $\alpha\approx 79\,^\circ$, both at the MIT and IMT.
$\alpha$ is the angle between the $y$-axis and the preferred direction  (see coordinates in Figure \ref{Fig1}c). 
By comparing two optical images taken at the same temperature ($T=168\,$K) in the heating branch (IMT) for two consecutive thermal cycles ({see Figure S3 in the supplement), we find that the overall domain geometry (domain size and orientation) for repeated measurements is reproducible, as deduced from the FFT of the images.
In real space, however, the domains may form in different locations in every heating cycle, despite a certain level of reproducibility.
Interestingly, the heating and cooling cycles show significant differences in domain geometry. These results show that while there is some domain pinning, the local transition temperature depends strongly on cycle-dependent domain nucleation and growth which changes the local strain distribution.
This indicates that the phase separation is not due to growth-induced local inhomogeneities (e.g.~in chemical composition) of the thin film, and that the domain configuration is plastic to a certain degree.

In Figure \ref{Fig1}b lower panel, we also include the reflectivity vs $T$ curve for a single pixel indicated by a red circle in the photomicrographs.
The discontinuous steps in reflectivity at the MIT and IMT, respectively, are indicative of a first-order phase transtition \cite{McLeod17}.

\subsubsection*{Spatially Resolved Transition Temperatures.}

\begin{figure}[t]
\includegraphics[width=\columnwidth]{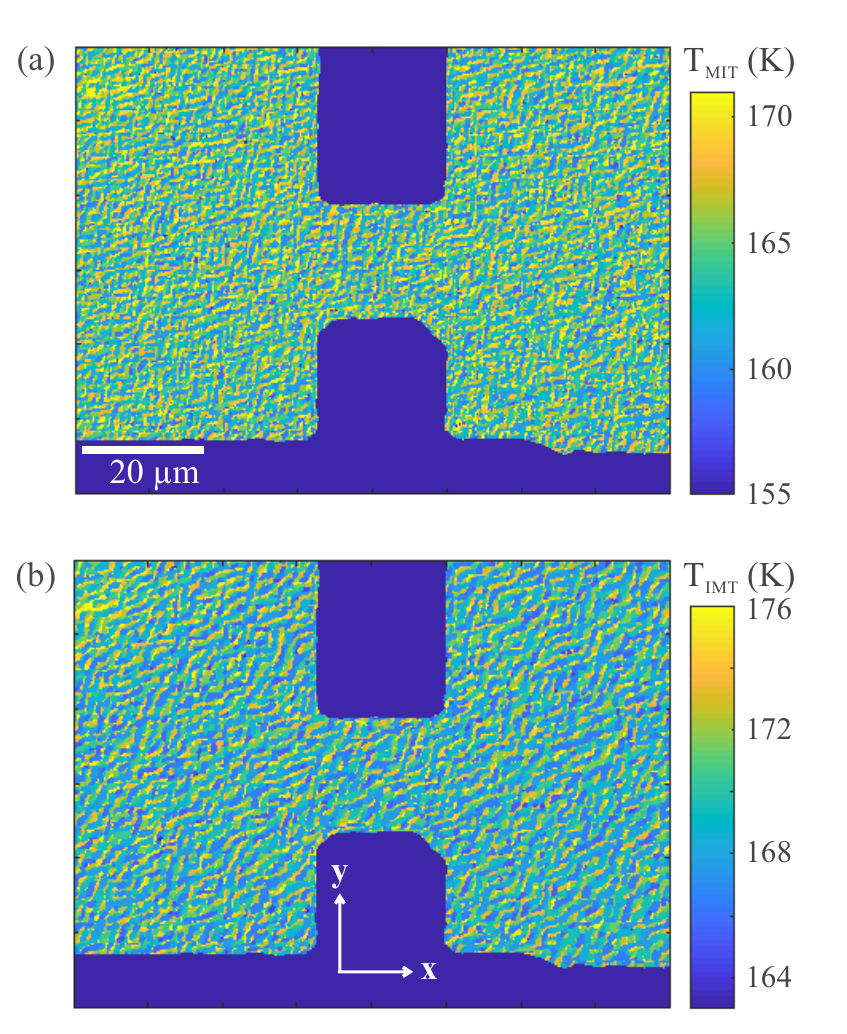} 
\caption{Local transition temperatures of V$_2$O$_3$ thin film device:
(a) $T_{\rm MIT}$ map, (b) $T_{\rm IMT}$ map.
Large dark blue areas indicate Au electrodes.}
\label{Fig2}
\end{figure}

By analyzing the steps in reflectivity for every pixel of the microscope image temperature series, maps of the local MIT and IMT temperatures $T_{\rm MIT}$ and $T_{\rm IMT}$, respectively, with a resolution of 1\,K can be derived (see Figure \ref{Fig2}).
The maps of $T_{\rm MIT}$ (Figure \ref{Fig2}a) and $T_{\rm IMT}$ (Figure \ref{Fig2}b) show different patches with different transition temperatures, whose shape and form resemble the herringbone domains in the photomicrographs.
We note that the transition temperature maps shown in Fig. 2 constitute a single instance of the local IMT/MIT temperature variation, representing the strain-induced domain structure, but not an actual deterministic transition temperature map.
These maps allow us in the numerical analysis (discussed below) to consider how the strain-induced variation in local MIT/IMT temperatures affects the electrothermal breakdown.

\subsection*{Current-Induced IMT -- Resistive Switching.}

In Figure \ref{Fig3}, we present electric transport data and simultaneously recorded photomicrographs for the current-induced IMT.
We also include the results of numerical simulations for direct comparison.
Note that we have added an animation to the supplementary information visualizing the resistive switching for every bias point.

\subsubsection*{Current-Voltage Characteristics.}

\begin{figure*}[!t]
\includegraphics{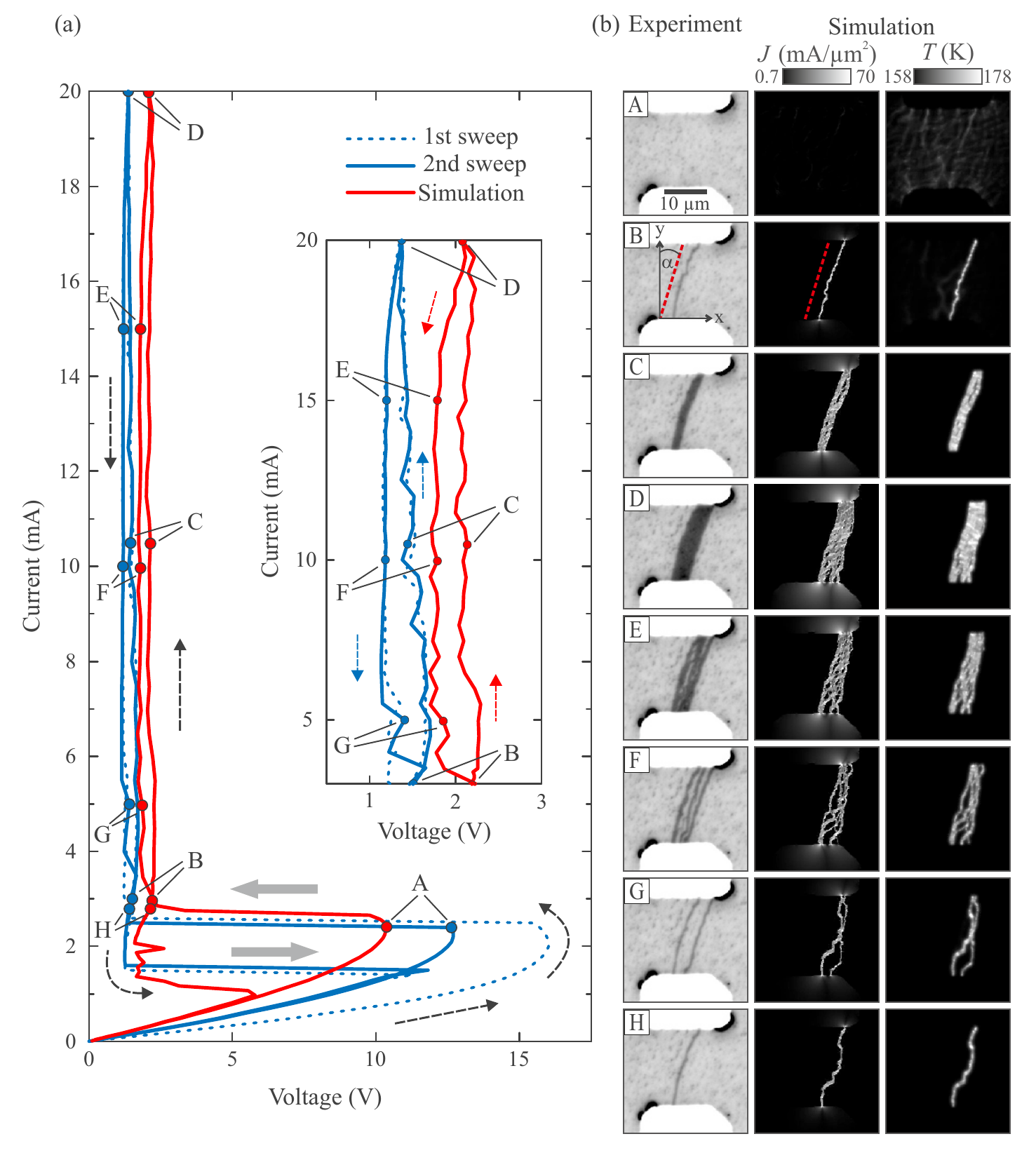} 
\caption{
Electrical breakdown in a planar V$_2$O$_3$ device at a base temperature $T_{\rm b}= 158\,$K at the onset of the IMT.
(a) Measured (blue) and simulated (red) IVCs.
Letters label data points for which images are shown in (b).
The arrows indicate the sweep directions.
The grey thick arrows indicate resistive switching to a low- and high-resistance state, respectively.
INSET: Zoom on the high-current section.
(b) LEFT COLUMN: Selection of photomicrographs acquired during the second current sweep.
MIDDLE AND RIGHT COLUMN: Selection of the simulated spatial distribution of current density $J$ and local device temperature $T$, respectively.
The dashed red lines for images B indicate a $\alpha\approx 17^\circ$ inclination of the filament with respect to the $y$-direction.}
\label{Fig3}
\end{figure*}

Before the measurement, the device was brought into the heating branch (at the base temperature $T_{\rm b}=158\,$K, cf.~Figure \ref{Fig1}b) via thermal cycling.
Then IVCs for two consecutive current sweeps were acquired (see  Figure \ref{Fig3}a).
During one sweep, the current increases from 0\,mA to the maximum value of 20\,mA and decreases back to 0\,mA.
The arrows in Figure \ref{Fig3}a indicate the sweep direction.
Starting at the origin of the graph, the IVCs progress almost linearly up to $I\approx 0.5\,$mA, which indicates an almost Ohmic resistance.
Upon further increasing $I$, the slope of the IVCs significantly increases, i.e., the differential resistance decreases.
A section with negative differential resistance follows.
A horizontal jump indicative of an instantaneous (within the measurement time scale) reduction of device resistance interrupts the IVC at $I\approx 2.5\,$mA.
In other words, an electrical breakdown of the device causes resistive switching to a low-resistance state. 
After the jump, the IVCs progress almost vertically with further increasing $I$.
When the current is reduced from its maximum value, the IVCs progress almost vertically down to $I\approx 1.8$\,mA, which is below the value of the first horizontal jump.
At this bias point, there is a second horizontal jump indicative of an instantaneous increase of device resistance, i.e., the device switches back to a high-resistance state.
A section with a decreasing slope and a linear section follows.

The IVCs of the first and second sweep are qualitatively similar. 
However, the maximum voltage, reached before the voltage jump appears in the upsweep, is reduced for the second sweep.
%
For sweeps following the second sweep (not shown), the maximum voltage does not change anymore and the IVCs are almost identical. 
Thermal cycling restores the IVC of the first sweep.
We infer from the reproducibility of the IVC measurements that the observed resistive switching is non-destructive.

In addition to the very pronounced hysteresis in the low-bias-current regime where resistive switching occurs, we observe a small hysteresis in the high-bias regime, i.e., the downsweep branches have lower voltages than the upsweep branches (see inset Figure \ref{Fig3}a). 
Furthermore, in this high-bias regime, the IVCs show small saw-tooth like jumps (see inset Figure \ref{Fig3}a).

\subsubsection*{Imaging Filaments.}

Simultaneously with recording IVCs during a current sweep, photomicrographs were acquired. 
We present a selection of these photomicrographs in Figure \ref{Fig3}b (left column) for the second current sweep. 
The animation attached as a supplement shows a movie of the full current sweep.

The following remarks are crucial for interpreting the results: 
As deduced from the images acquired during temperature sweeps, metallic and insulating domains have different optical reflectivity, allowing us to assign metallic and insulating resistivities to dark and bright regions of the device, respectively. 
From Figure \ref{Fig1}c, we learn that dark areas in the photomicrograph indicate a metallic V$_2$O$_3$ phase, while bright areas indicate an insulating phase.

Photomicrograph A in Figure \ref{Fig3}b was acquired at bias point A on the upsweep branch of the second current sweep, just before the large voltage jump appears. 
The photomicrographs acquired between zero bias and bias point A are all identical (see animation in the supplement). 
The V$_2$O$_3$ thin film between the electrodes is insulating (bright) with several small metallic (dark) inclusions, and the device is in a high-resistance state.  
The metallic inclusions appeared after the voltage jump of the first current sweep.
Photomicrograph B corresponds to bias point B immediately after the large voltage jump, indicating an abrupt switching of the device to a low-resistance state.
A narrow metallic (dark) filament connects the electrodes at an oblique angle $\alpha$, shunting the insulating film.  
With further increasing current, the filament width increases (see photomicrographs B--D) by growing to the right.
Bias point D is at $I=20\,$mA, which terminates the upsweep branch.
When the current is reduced, the filament splits into several branches (photomicrograph E). 
When the current is reduced further, the splitting results in multiple parallel filaments (photomicrograph F). 
Then these filaments disappear one after the other (photomicrographs F-H). 
Photomicrographs H and B are taken at almost the same current value and are almost identical. 
Finally, when the current is reduced to a bias-point just below the second voltage-jump, indicating an abrupt increase of the device resistance, the last remaining filament disappears, and the device returns to an insulating, high-resistance state (see animation in the supplement).

\subsubsection*{Discussion of Experimental Results.} %

\begin{figure}[!b]
\includegraphics[width=\columnwidth]{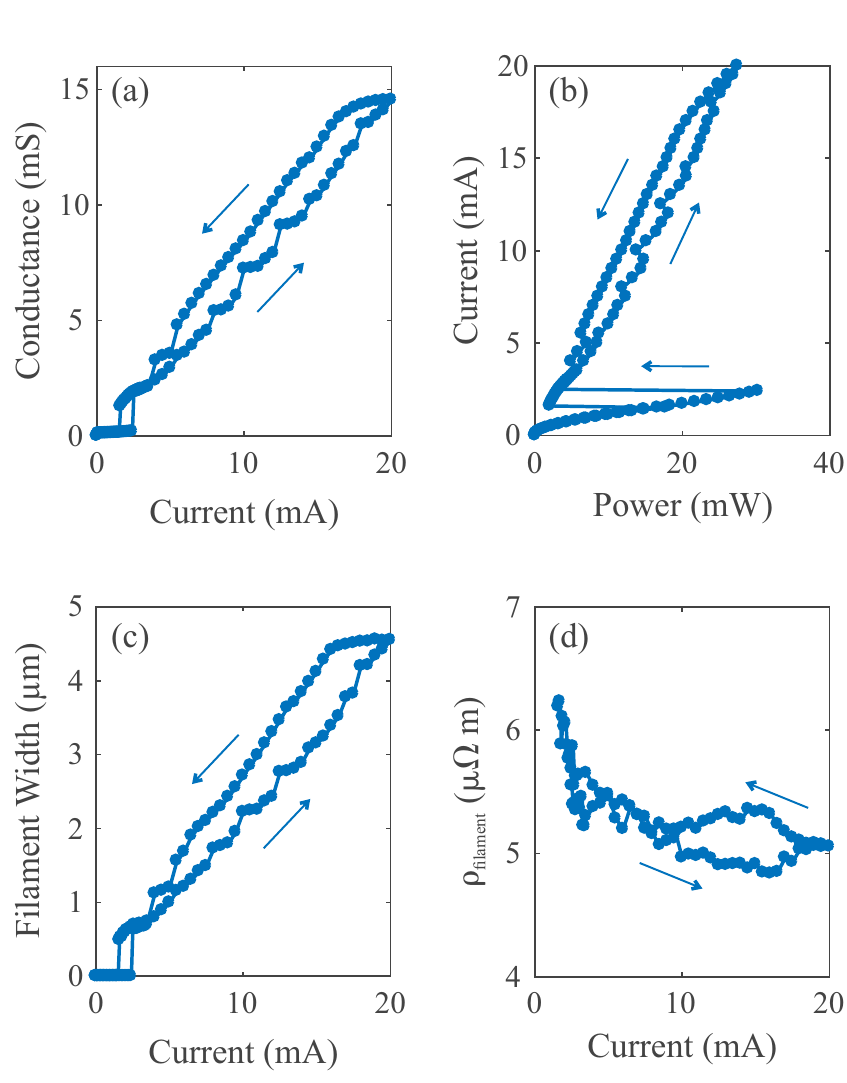}
\caption{
Relations derived from the IVC and photomicrographs of the second current sweep in Figure \ref{Fig3}.
(a) conductance vs.~current, (b) current vs.~power, (c) estimated total filament width vs.~current, and (d) resistivity of the metallic phase estimated from filament width vs.~current.}
\label{Fig4}
\end{figure}

We derived additional relations (Figure \ref{Fig4}) from the IVC and photomicrographs of the second current sweep.
These relations help to interpret the results. 
The derivation of the conductance ($G=I/V$) vs.~current and current vs.~power ($P=I\cdot V$) relations (Figure \ref{Fig4}a, b, respectively) is straightforward. 
In Figure \ref{Fig4}c, the filament width was estimated by dividing the dark filament area in the photomicrographs by the filament length, which was estimated by approximating the filament as a straight line. 
Similarly, the filament resistivity $\rho_{\rm filament}$ was derived, assuming that the filament extends throughout the thickness of the V$_2$O$_3$ film.

Now, we revisit the IVCs and the photomicrographs shown in Figure \ref{Fig3}.
The overall behavior of the device is characteristic of an electrothermal breakdown \cite{Gurevich87}.
The increasing slope, after the linear (Ohmic) section in the low-bias-current regime before switching to the high-resistance state, can be explained by a decrease of the device resistance  due to Joule heating.
According to the $R(T)$ curve shown in Figure \ref{Fig1}b, close to the IMT an increase in device temperature results in a considerable decrease in device resistance.
This effect produces the observed back-bending of the IVC. 
The system then becomes electrothermally unstable, and the horizontal jump is the result of the formation of an electrothermal filament.
Figure \ref{Fig3}b image B clearly shows the appearance of this metallic filament.
The vertical progression of the IVC with increasing $I$ after switching to the low-resistance state is associated with an increase in filament width (see Figure \ref{Fig3}b images B--D).
The second horizontal jump during the return current sweep indicates an abrupt increase of the device resistance associated with the disappearance of the conducting filament (see animation in supplement).

We note that, if the resistive switching were purely electric-field-driven, the resistive switching would occur at maximum voltage. 
However, we do not observe resistive switching at a maximum voltage (see Figure \ref{Fig3}a) but at a maximum power (see Figure \ref{Fig4}b), which is further evidence for an electrothermal breakdown.
According to Figure \ref{Fig4}, in the high-bias regime after resistive switching, the device conductance, power, and filament width have an overall linear current dependence, while the filament resistivity is almost constant.
Again, these dependencies are characteristic of an electrothermal breakdown, where the device current flows in a metallic filament, and the current controls the width of this filament via Joule heating.
We note that the initial steep decrease of $\rho_{\rm filament}$ vs $I$ in Figure \ref{Fig4}d might be due to the possiblity that right after its formation (or before its extinction), the filament does not yet expand throughout the entire film thickness and/or that the filament width is below the spatial resolution of the microscope.
In both cases, we would slightly overestimate $\rho_{\rm filament}$.
However, we also note that the independently determined conductance vs current (Figure \ref{Fig4}a; from IVCs) and filament width vs current (Figure \ref{Fig4}c; from images) curves match very closely, including the jumps at disctinct current values.
This indicates that the minimum filament width is very likely not much smaller than what we observe by imaging.
Moreover, this again clearly demonstrates how the overall device resistance in the low-resistance state is governed by the optically detected filamentary structures.

In addition to all the effects that are typical for an electrothermal breakdown, we observe atypical effects, which can only be explained by effects that are not included in the most simple electrothermal model.
First, the filament is not straight along the $y$-axis, but rather connects the electrodes at an oblique angle $\alpha$ of approximately $17^\circ$ with respect to the $y$-direction (see image B in Figure \ref{Fig3}b).
In a homogeneous medium, one would expect that the shortest possible current path is preferred, and the filament should follow the $y$-direction.
Second, during the downsweep the filament first splits into several branches, and then it divides into multiple parallel filaments (see images E and F in Figure \ref{Fig3}b). 
Because adjacent regions in the film are thermally coupled in the lateral direction, for a homogeneous medium one would expect that as current decreases the filament shrinks from its sides, since the filament temperature is highest in its center, so that the colder sides would transition into the metallic state more readily.
Third, for high bias currents in the low-resistance state, there is hysteresis in the IVC (see inset Figure \ref{Fig3}a) and in the conductance, power and filament width vs.~current relations (Figure \ref{Fig4}). 
However, after an electrothermal breakdown in a homogeneous medium without thermal hysteresis, the current should control these parameters without hysteresis in the IVC.

\subsubsection*{Comparison of Simulations with Experimental Results.} %

We developed a numerical model to investigate whether the electrical breakdown is electrothermal and whether the phenomena which are atypical for an electrothermal breakdown in a homogeneous medium without thermal hysteresis result from the strain-induced spatial variation of the MIT/IMT temperatures (see Figure \ref{Fig2}).

The 2D resistor network model includes the experimentally determined spatial distribution of $T_{\rm MIT}$ and $T_{\rm IMT}$ (cf.~Figure \ref{Fig2}) and allows us to calculate IVCs and the spatial distribtion of temperature $T$ and current density $J$.
For details, see the Methods section and the supplement.

The simulation results are included in Figure \ref{Fig3} for direct comparison with the experimental data.
The model was numerically stable over almost the whole parameter space, except for a section in the first current sweep after the resistive switching.
For this reason, we focus our discussion on the second sweep.
We note that the simulation of the first sweep reveals that after return to zero current, some resistors remain in the metallic state.
This is consistent with the imaging results, which show that metallic inclusions appear after the first sweep.
For the simulation of the second sweep, this information from the simulation of the first sweep has been included, which explains why we find different results also in the simulations of different sweeps.

The heuristic numerical model, which includes strain-effects in the V$_2$O$_3$ thin film indirectly via the IMT/MIT temperature maps, reproduces the resistive switching very well.
Comparing the simulated IVC with the measured IVC in Figure \ref{Fig3}a, the threshold currents for resistive switching are very similar, and the maximum voltage in the simulation is only 14\,\% smaller than in the experiment. 
The model reproduces the appearance and disappearance of the metallic filament accompanying resistive switching (see Figure \ref{Fig3}b and animation in the supplement).
Additionally, the effects, which are atypical for an electrothermal breakdown (discussed above), are reproduced as well.

The agreement of the electrothermal model with the experimental results supports the hypothesis that an electrothermal breakdown induces resistive switching.  
Simulated images for bias points A and B in Figure \ref{Fig3}b demonstrate the current- and temperature redistribution.
As discussed in the introduction, this redistribution results from a runaway effect driven by electrothermal instability.
In particular, the simulation for bias point A shows already an inhomogeneous $T$ distribution within the entire device area prior to resistive switching.
After switching to the low-resistance state at bias point B, a single metallic filament appears in the optical image, and the simulation shows that the current density is concentrated in this filament, with a local temperature of $\sim 20\,$K above the base temperature $T_{\rm b}$, while the area outside the filament has cooled down.
In bias point B, the simulation still shows a few filamentary regions with slightly enhcanced temperature, while with further sweeping up the bias current, these areas also cooled down close to $T_{\rm b}$.

Remarkably, the simulated data do not only reproduce all the experimentally observed features in the IVCs, but also in the shape and direction of the metallic filament.
The filaments in the photomicrographs and the simulation both connect the electrodes at an oblique angle of $\alpha\approx 17\,^\circ$, following roughly one preferred axis of the herringbone domains (see Figure \ref{Fig1}c).
The minimum filament width, right before the device switches back to the high-resistance state, is approximately $0.5\,\mu$m, which corresponds to the domain size.
However, we note that the value of $0.5\,\mu$m has to be considered as an upper limit, as this is also the spatial resolution of the optical microscope.
These observations support the idea that the herringbone domain structure plays an essential role in resistive switching.

The effects atypical of an electrothermal breakdown, like hysteresis and sawtooth-like jumps in the IVC, filament branching and splitting seem to have the same cause: 
a hysteretic and strain-induced heterogenous electrothermal medium.

Considering the overall shape of the IVCs shown in Figure \ref{Fig3}, the voltage is almost constant in the low-resistance state.
This behavior is typical for an electrothermal breakdown, where a Joule heating filament controls the voltage across the device.
Usually, the filament growth with applied current is reversible, and the increasing and decreasing current curves are congruent in this bias regime.
However, the IVCs are hysteretic (see inset in Figure \ref{Fig3}a).
We argue that the hysteresis in the thermally driven MIT/IMT, i.e., the shift between the MIT- and IMT-temperatures in V$_2$O$_3$, is causing this hysteresis in the IVCs.

The sawtooth-like IVC jumps (inset in Figure \ref{Fig3}a) are due to discontinuous generation and extinction of entire filaments or segments of filaments. We attribute this to filament edge pinning of boundaries between the insulating and metallic phase due to the spatial variation in local IMT-temperatures.
In particular, the filament does not grow continuously as the current is increased but switches adjacent domains to the metallic state as soon as the dissipated power is sufficient.
For increasing currents, only a single filament is visible in the photomicrographs, whose width increases for increasing currents.
For decreasing currents, the expected behaviour would be that the filament width decreases continuously.
However, the filament branches and splits into multiple filaments (see Figure \ref{Fig3} and the animation attached as a supplement).
The varying local MIT/IMT-temperatures (see Figure \ref{Fig2}) define percolation paths to which the filaments can lock in.
Consequently, when the current is reduced, it distributes along these percolation paths.
The geometry of these paths closely resembles the sponteneously forming domain configuration observed during temperature sweeps in the phase coexistence regime at the MIT.
For the same reason, the filament is not connecting the electrodes along the shortest path but at an oblique angle.
In contrast to the filament in the photomicrographs, the filament in the simulation consists of multiple filaments for biasing on the increasing current branch, although it is more compact than for the decreasing current branch. 
One explanation is that we cannot resolve these multiple filaments in the photomicrographs due to the limited resolution of the microscope. 
Another explanation is that the simulation uses transition temperature maps derived from a series of fixed domain patterns, while in reality the domains are plastic (see discussion of "Phase Separation at the MIT/IMT" and Figure S3 in the supplement), i.e., domain walls can move.
Due to this plasticity, there might be a tendency of the domains to merge, which is not reflected in the numerical model.

\section*{Conclusion}
\label{sec:Conclusion}

We demonstrated electrothermally induced resistive switching in a current-biased planar V$_2$O$_3$ thin-film device with a base temperature at the onset of the IMT. 
Photomicrographs show the appearance of a metallic filament that accompanies resistive switching. 
A model of the resistive switching reveals that the spatial variation in local MIT/IMT-temperatures influences the details of the electrothermal breakdown.

The variation of local MIT/IMT temperatures is the result of strain minimization in the V$_2$O$_3$ thin film, and we conclude that elastic energy influences resistive switching properties by affecting the filament configuration.
Therefore, the influence of strain on the dynamics of switching devices should be taken into account and may allow for novel ways of tuning their properties.

These findings demonstrate the crucial role played by Joule heating and strain-effects in this class of memristive devices, which are considered as promising building blocks in neuromorphic computing.
Obviously, the approach taken in this work can also be applied to other Mott insulator systems.

\section*{Methods}
\label{sec:Methods}

\textbf{Device fabrication:} The 300-nm-thick V$_2$O$_3$ thin film was grown by rf magnetron sputtering of a V$_2$O$_3$ target on a r-cut sapphire substrate; see Ref.~\onlinecite{Stewart12} for details.
Subsequently, by an optical lithography lift-off process, we prepared several-100-nm-thick Au contact pads on top of the unpatterned  V$_2$O$_3$ film, to define 12 devices on the chip.
Each device consists of a 19.5-$\mu$m-long gap between two 20-$\mu$m-wide Au electrodes.

\textbf{Experimental set-up:} We used a wide-field optical microscope \cite{Lange17} with the device mounted in vacuum, in a continuous He gas-flow cryostat, and electrically connected to perform electric transport measurements, simultaneously with the imaging of the device.
The microscope has a spatial resolution of about $0.5\,\mu$m.
The illumination is monochromatic with a 528\,nm wavelength, and the field of view is $500\,\mu{\rm m} \times 500\,\mu$m.

Optical microscopy allows for imaging the phase separation in the V$_2$O$_3$ thin film, because the metallic- and insulating-phase have different reflectivity (see Figure \ref{Fig1}c).
Note that counterintuitively the insulating phase has a higher reflectivity than the metallic phase.
The reason for this is that the monochromatic illumination at 528\,nm is above the plasma frequency for both phases, and therefore the reflectivity contrast is not caused by the concentration of free charge carriers.
Instead, e$_g^\pi$  to a$_{\rm 1g}$ (lower Hubbard band) \cite{Poteryaev07,Mo03} interband transitions dominate the insulating-phase reflectivity, while transitions from the quasi-particle peak to the a$_{\rm 1g}$ upper Hubbard band \cite{Stewart12} dominate the metallic-phase reflectivity.

We measured the electric transport properties in a two-point configuration with a Keithley 2400 source/meter unit configured as current-source.
IVCs were taken from 0 to 20\,mA and back to 0 (between 0 and 3.5\,mA the step size was 0.1\,mA, then the step size was increased to 0.5\,mA, and between 3.0 and 0\,mA the step size was reduced to 0.1\,mA).

The \textbf{numerical model} is based on a two-dimensional resistor network, for which the current distribution is solved via mesh current analysis, while the thermal properties are modeled with a finite difference approximation of the time-dependent heat equation. 
The heat conduction was incorporated using the backward Euler method.
Latent heat was included.
We accounted for the thermal coupling to the cryostat cold-plate by estimating the sapphire substrate's thermal conductivity.
The model was numerically stable over almost the whole parameter space, except for a small section in the first current sweep after the resistive switching.
A detailed description, based on Ref.~\onlinecite{Lange18}, is given in the supplement.

Every resistor of the 2D network has an individual hysteretic $R(T)$ dependence with MIT/IMT temperatures corresponding to the local values of Figure \ref{Fig2}. 
This way, the strain-induced variation of the MIT/IMT temperatures is included in the model. 
Prior to the simulation of an entire current-sweep sequence, for each pixel the high- or low-resistance state within the hysteretic $T$-range is set, depending on history.
This procedure takes into account, e.g., previous current-sweep sequences and hence can capture history effects in simulations for different current sweeps.
For a given bias current value, the mesh current analysis and the time-dependent heat equation is solved with small time steps until the solution reaches a steady state.
Subsequently, the procedure is repeated with an incremented current value. 
This way, we calculate the IVC and the spatial distribution of the current density $J$ and the temperature $T$.

\section*{Associated content}

\textbf{Supporting Information}

Measured IVC, photomicrographs, and simulation data of the electrical breakdown in a planar V$_2$O$_3$ device combined in one animation (AVI).

Further details on the phase separation at the MIT/IMT (Evolution of domain configurations across the MIT/IMT, directionality and periodicity of the domain patterns, reproducibility of the domain configuration).
Description of the numerical model (resistor network and mesh current analysis, definition of temperature-dependent resistivity, finite difference approximation to heat equation, simulation of electric breakdown in V$_2$O$_3$) (PDF).

\textbf{Author Contributions}

M.L.~did the measurements with the help of T.L.~and D.S..
M.L.~programmed the numerical model.
S.G.~conceived and coordinated the study.
Y.K.~and N.M.V.~fabricated the sample.
M.L.~and S.G.~developed the cryogenic microscope.
S.G.~wrote the manuscript with inputs from M.L., Y.K., I.K.S, R.K.~and D.K..
All authors discussed the results.
All authors have given approval to the final version of the manuscript.

‡ These authors  (M.L. and S.G.) contributed equally.

\textbf{Notes}

The authors declare no competing financial interests.

\section*{Acknowledgments}

This work was supported as part of the Quantum Materials for Energy Efficient Neuromorphic Computing (Q-MEEN-C) Energy Frontier Research Center (EFRC), funded by the U.S. Department of Energy, Office of Science, Basic Energy Sciences under Award \# DE-SC0019273.
Part of the fabrication process was done at the San Diego Nanotechnology Infrastructure (SDNI) of UCSD, a member of the National Nanotechnology Coordinated Infrastructure (NNCI), which is supported by the National Science Foundation under grant ECCS-1542148.
Yoav Kalcheim acknowledges the support of a Norman Seiden Fellowship in Nanotechnology and Optoelectronics.

\bibliography{V2O3microscopy}

\end{document}


\title{Supplementary Information: Optical imaging of strain-mediated phase coexistence during electrothermal switching in a Mott insulator}
\author{Matthias Lange}
\affiliation{%
Physikalisches Institut, Center for Quantum Science (CQ) and LISA$^+$,
Eberhard Karls Universit\"at T\"ubingen,
Auf der Morgenstelle 14,
72076 T\"ubingen, Germany}

\author{Stefan Gu\'enon}
\email{stefan.guenon@pit.uni-tuebingen.de}
\affiliation{%
Physikalisches Institut, Center for Quantum Science (CQ) and LISA$^+$,
Eberhard Karls Universit\"at T\"ubingen,
Auf der Morgenstelle 14,
72076 T\"ubingen, Germany}

\author{Yoav Kalchheim}
\affiliation{%
Center for Advanced Nanoscience, Department of Physics, 
University of California –- San Diego, 
9500 Gilman Drive, La Jolla, CA92093-0319, USA}
\affiliation{%
Department of Materials Science and Engineering, Technion -- Israel Institute of Technology, Technion City, 32000 Haifa, Israel}

\author{Theodor Luibrand}
\affiliation{%
Physikalisches Institut, Center for Quantum Science (CQ) and LISA$^+$,
Eberhard Karls Universit\"at T\"ubingen,
Auf der Morgenstelle 14,
72076 T\"ubingen, Germany}

\author{Nicolas Manuel Vargas}
\affiliation{%
Center for Advanced Nanoscience, Department of Physics, 
University of California –- San Diego, 
9500 Gilman Drive, La Jolla, CA92093-0319, USA}

\author{Dennis Schwebius}
\affiliation{%
Physikalisches Institut, Center for Quantum Science (CQ) and LISA$^+$,
Eberhard Karls Universit\"at T\"ubingen,
Auf der Morgenstelle 14,
72076 T\"ubingen, Germany}

\author{Reinhold Kleiner}
\affiliation{%
Physikalisches Institut, Center for Quantum Science (CQ) and LISA$^+$,
Eberhard Karls Universit\"at T\"ubingen,
Auf der Morgenstelle 14,
72076 T\"ubingen, Germany}

\author{Ivan K.~Schuller}
\affiliation{%
Center for Advanced Nanoscience, Department of Physics, 
University of California –- San Diego, 
9500 Gilman Drive, La Jolla, CA92093-0319, USA}

\author{Dieter Koelle}
\affiliation{%
Physikalisches Institut, Center for Quantum Science (CQ) and LISA$^+$,
Eberhard Karls Universit\"at T\"ubingen,
Auf der Morgenstelle 14,
72076 T\"ubingen, Germany}


\maketitle

\makeatletter 
\renewcommand{\thefigure}{S\arabic{figure}}

\section{Phase Separation at the MIT/IMT}

This section provides further details on the phase separation of the investigated V$_2$O$_2$ film across the metal-to-insulator transition (MIT) and the insulator-to-metal transition (IMT).

\subsection{Evolution of domain configurations across the MIT/IMT}

\begin{figure}[b!]
\includegraphics[width=\columnwidth]{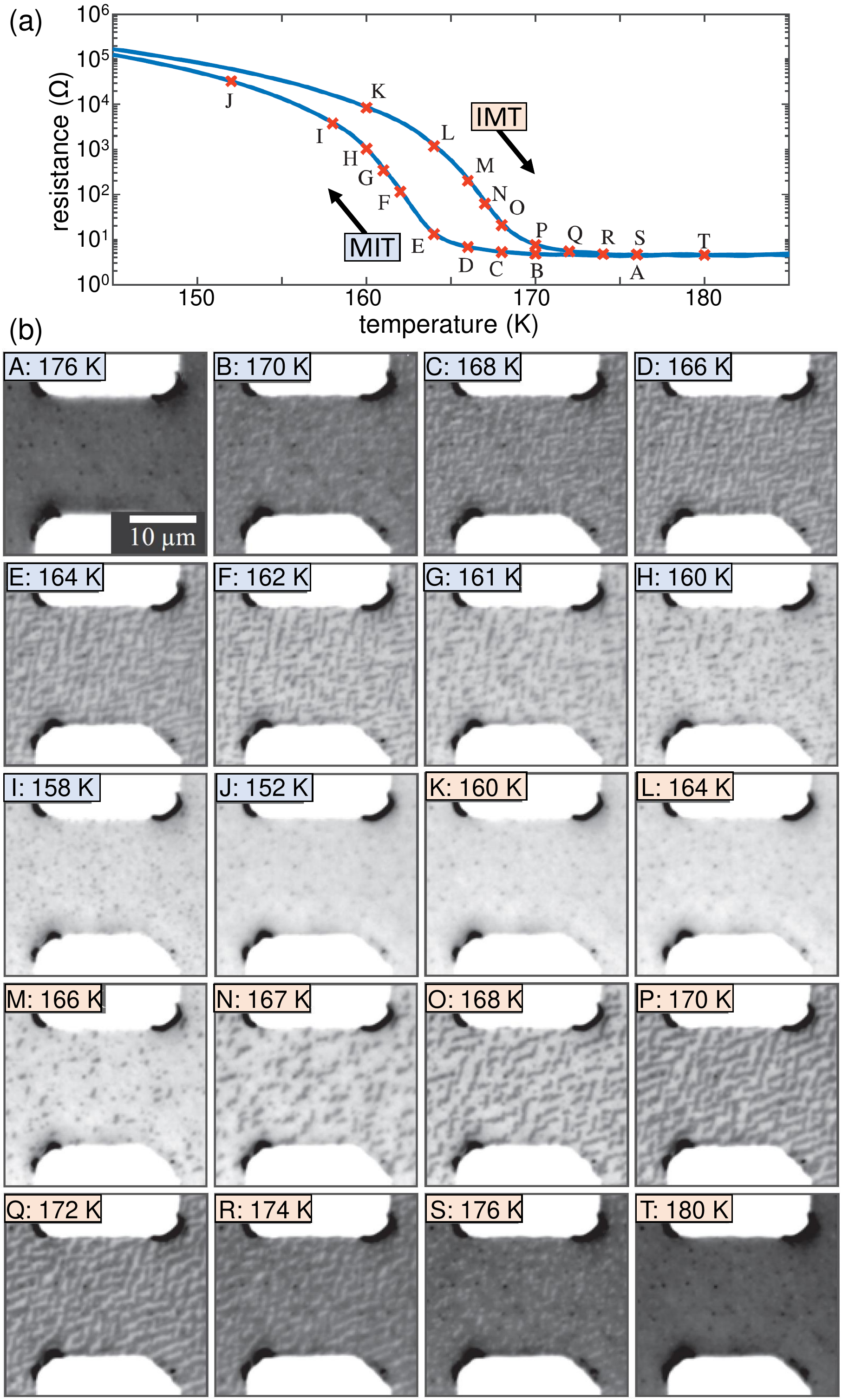}
\caption{
Phase separation in V$_2$O$_3$ at the MIT/IMT:
(a) $R(T)$ curves.
(b) Images acquired at the points, labeled A to T in (a), during cool-down and warm-up of the sample.
\label{fig:phasesep}}
\end{figure}

Figure~\ref{fig:phasesep}(b) shows a more detailed series of photomicrographs at temperatures $T$ across the MIT/IMT. 
The resistance $R$ vs $T$ curves (Fig.~\ref{fig:phasesep} (a)) were acquired with a bias current $I=5\,\mu$A. 
The capital letters indicate the acquisition temperatures of the photomicrograph series.
The $R(T)$ curves show a four-order of magnitude change of $R$ across the MIT/IMT, with a thermal hysteresis of $\approx 5\,$K between cooling and heating branches.
The minimum temperature within the thermal cycle was 80\,K.

The transition between the two phases progresses through the formation of domain patterns with dark and bright contrast that are identified as metallic and insulating regions, respectively.
Starting from a homogeneous metallic state (A: 176\,K), small insulating islands begin to form (B: 170\,K) that grow and connect to form a herringbone-like domain pattern (C: 168\,K and D: 166\,K). 
At this point, the metallic domains still provide continuous paths connecting the electrodes. 
These get disconnected when the temperature is further decreased (E: 164\,K and F: 162\,K), which goes along with a steep increase in resistance, as the current now has to flow through parts of the film that are in the insulating phase. 
At even lower temperatures (G: 161\,K, H: 160\,K, and I: 158\,K) the V$_2$O$_3$ film is in a state where metallic patches are embedded within an insulating matrix. 
The size of the metallic domains decreases with decreasing temperature until the sample is in an almost  homogeneous insulating state (J: 152\,K).

\begin{figure*}[t!]
\includegraphics[width=13cm]{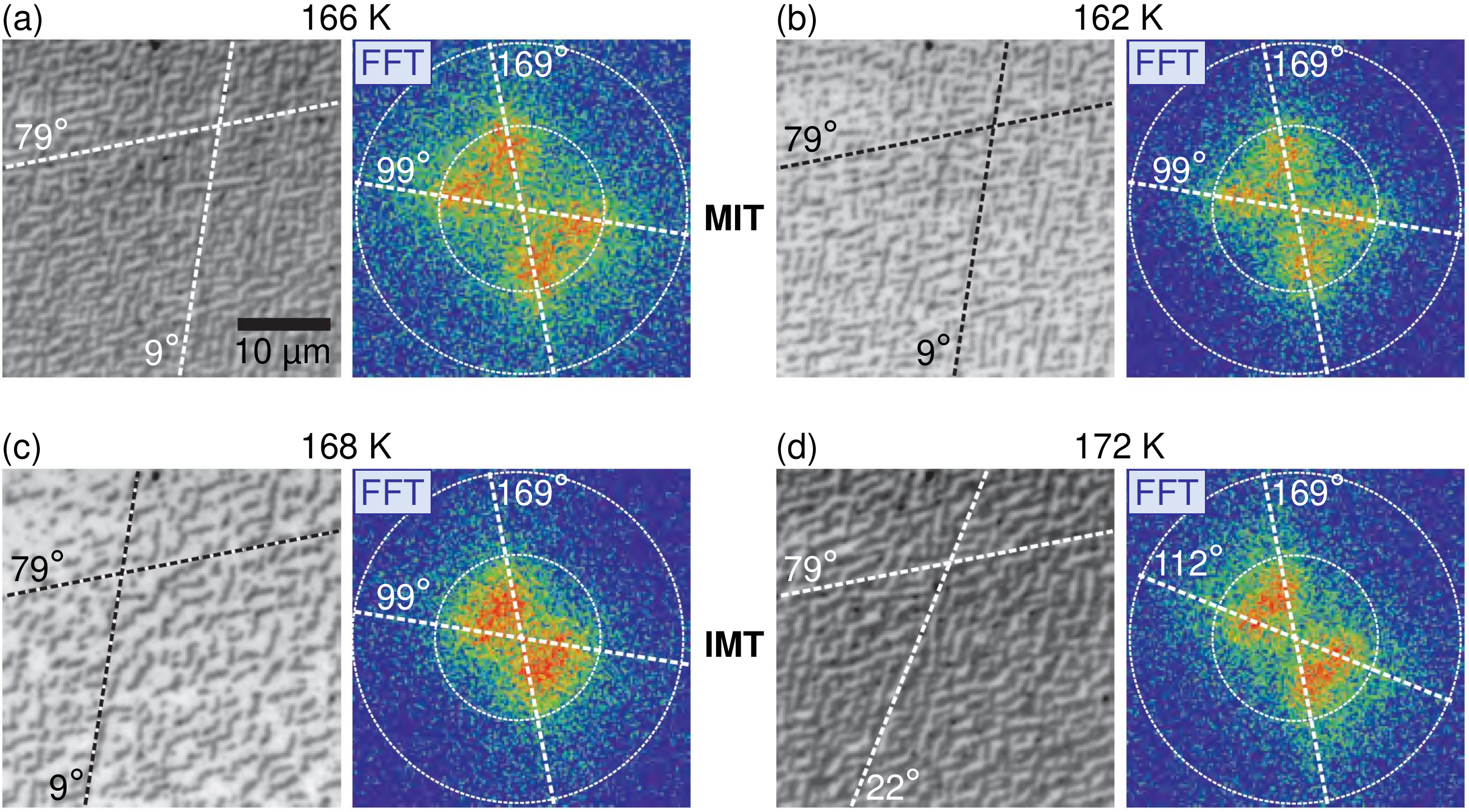}
\caption{
Domain configuration (left) and fast Fourier transform (right) for selected temperatures of (a) $166\,\mathrm{K}$ and (b) $162\,\mathrm{K}$ in the cooling branch (MIT) and (c) $168\,\mathrm{K}$ and (d) $172\,\mathrm{K}$ in the heating branch (IMT).
The dashed lines highlight the predominant directionality of the long axis of domains in the real-space images and the directions at which the peaks occur in the FFT.
The angles are measured clockwise with respect to the vertical axis.
The inner (outer) circles in the FFT correspond to a spatial frequency of $1/\mathrm{\mu{m}}$ ($2/\mathrm{\mu{m}}$).
\label{fig:phaseFFT}}.
\end{figure*}

The heating branch shows a slightly different behavior.
Beginning from the insulating state (K: 160\,K), small metallic islands appear (L: 164\,K and M: 166\,K) that connect to elongated domains (N: 167\,K and O: 168\,K).  
These domains are wider than the domains in the cooling branch and have a different preferred direction. 
Upon further increase of the temperature, more metallic domains appear and long metallic paths are created that eventually form continuous metallic paths that connect the electrodes (P: 170\,K). 
The metallic regions grow (Q: 172\,K and R: 174\,K) until only isolated insulating patches are left (S: 176\,K). 
Eventually the film arrives in a homogeneous metallic state (T: 180\,K).

\subsection{Directionality and periodicity of the domain patterns}

The preferred direction of the domains and their periodicity can be estimated from the fast Fourier transform (FFT) of the optical images.
This is shown for selected temperatures in Fig.~\ref{fig:phaseFFT}. 
Angles are measured clockwise with respect to the vertical axis, i.e., those correspond to the definition of $\alpha$ for optical images in the manuscript.
The FFT of the domain configuration in the MIT [Fig.~\ref{fig:phaseFFT}~(a) and (b)] clearly shows four peaks that correspond to directions of $\approx 99 \,\mathrm{^\circ}$ and  $\approx 169 \,\mathrm{^\circ}$, with a mean periodicity of $\approx 1.3\,\mathrm{\mu{m}}$ at $166\,\mathrm{K}$ and $\approx 1.5\,\mathrm{\mu{m}}$ at $162\,\mathrm{K}$.  
The preferred orientations of the long axis of the domains in real space are perpendicular to the directions found in the FFT and are indicated by the broken lines at  $\approx 9 \,\mathrm{^\circ}$ and  $\approx 79 \,\mathrm{^\circ}$ in the optical images. 
Figure~\ref{fig:phaseFFT}~(c) and (d) show the domain configuration and FFT for temperatures of $168\,\mathrm{K}$ and $172\,\mathrm{K}$ in the IMT. 
At $T=168\,$K, it is not possible to distinguish the peaks of the $99\,\mathrm{^\circ}$ and $169\,\mathrm{^\circ}$-axes in the FFT. 
However, the real space image at $168\,\mathrm{K}$ shows that many of the domains still are oriented along the $9\,\mathrm{^\circ}$ and $79\,\mathrm{^\circ}$-axes, while some domains have an orientation that lies in between these axes. 
At $172\,\mathrm{K}$, on the other hand, only few domains are oriented along the $9\,\mathrm{^\circ}$-axis, and the preferred directions become $22\,\mathrm{^\circ}$ and $79\,\mathrm{^\circ}$, which is observed in the FFT as a narrowing of the peaks in angular direction.
The mean periodicity of the domains in the heating branch (IMT) is found to be $\approx 2.2\,\mathrm{\mu{m}}$ at $168\,\mathrm{K}$ and $\approx 1.8\,\mathrm{\mu{m}}$ at $172\,\mathrm{K}$.

\subsection{Reproducibility of the domain configuration}

\begin{figure}[t!]
\includegraphics[width=0.8\columnwidth]{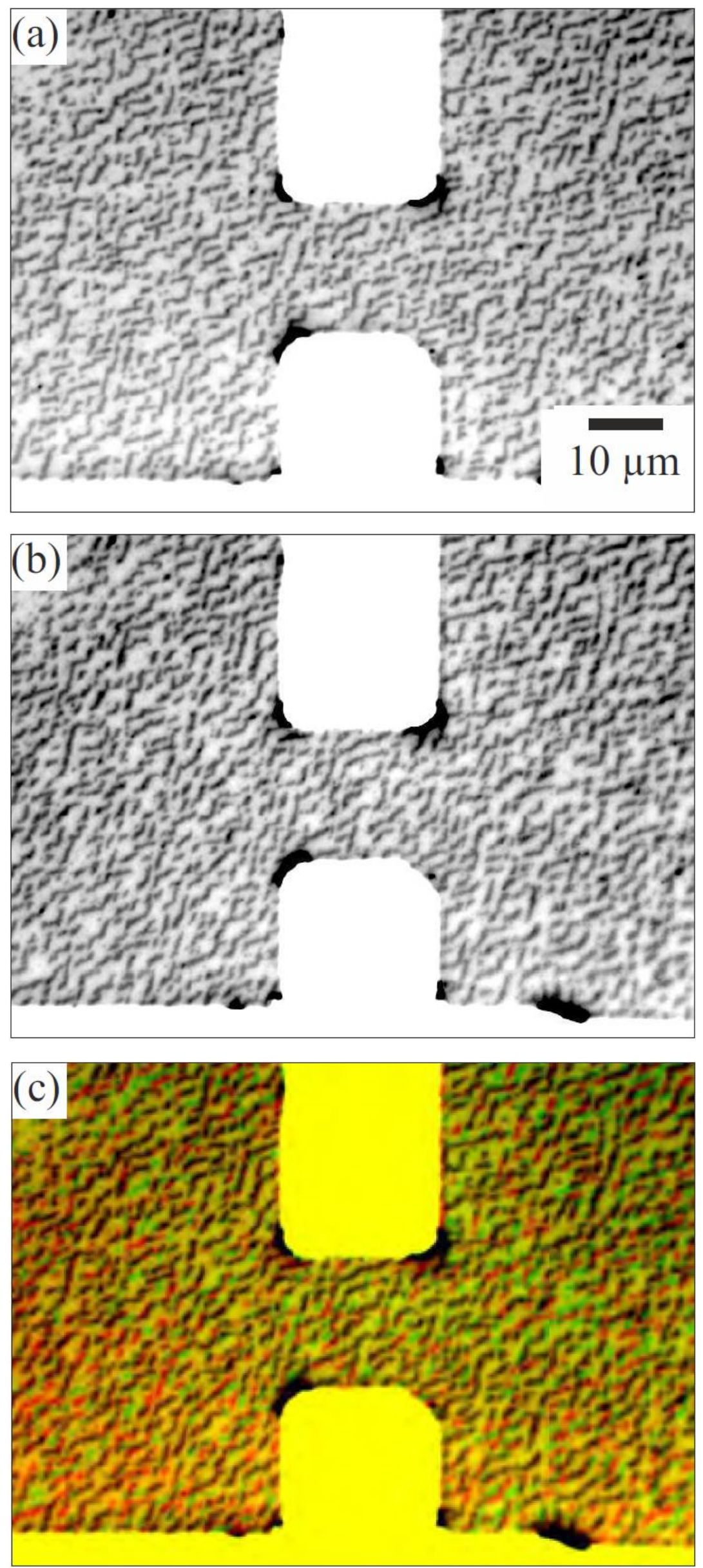}
\caption{
Reproducibility of domain configuration.
(a),  (b) Optical images, recorded at 168\,K in the heating branch for two subsequent runs with a full thermal cycle in between.
(c) Color-coded superposition of images from (a) and (b).
Metallic domains that are present in both (a) and (b) appear dark.
Metallic domains only present in (a) are green, and only present in (b) are red.
Insulating domains present in both micrographs are shown in yellow.
\label{fig:domain_repeatability}}
\end{figure}

Figure~\ref{fig:domain_repeatability}~(a) and (b) show the domain configuration at $168\,\mathrm{K}$ in the heating branch for two subsequent measurement runs. 
A complete thermal cycle (heating to room temperature and cooling to 80\,K) separates the two images. 
While the overall density of metallic domains is similar, the domain patterns differ significantly. 
For comparison, a color-coded superposition of the images in Fig.~\ref{fig:domain_repeatability}~(a),(b) is shown in Fig.~\ref{fig:domain_repeatability}~(c). 
Metallic domains that are present in both, Fig.~\ref{fig:domain_repeatability}~(a) and (b), appear dark.
Domains that are present only in image (a) are represented in green, those that are only present in (b) in red. 
It is apparent, that the domain configuration for repeated measurements is neither deterministic nor random.
The same holds true for other temperatures in both the MIT and IMT. 
This indicates that the phase separation is not due to growth-induced local inhomogeneities of the film, for example in chemical composition, that could lead to a spatially varying transition temperature but rather is caused through minimization of local strain as proposed by McLeod~\textit{et al.}~\cite{McLeod17}.

\section{Numerical Model}

This section describes the numerical model, which has been used to simulate the current-voltage characteristics (IVCs) and spatial distribution of temperature $T$ and current density $J$ in the V$_2$O$_3$ films.
This description is based on chapter 5.4 in Ref.~\onlinecite{Lange18}.

\subsection{Resistor network and mesh current analysis}

\begin{figure}[b!]
\includegraphics[width=\columnwidth]{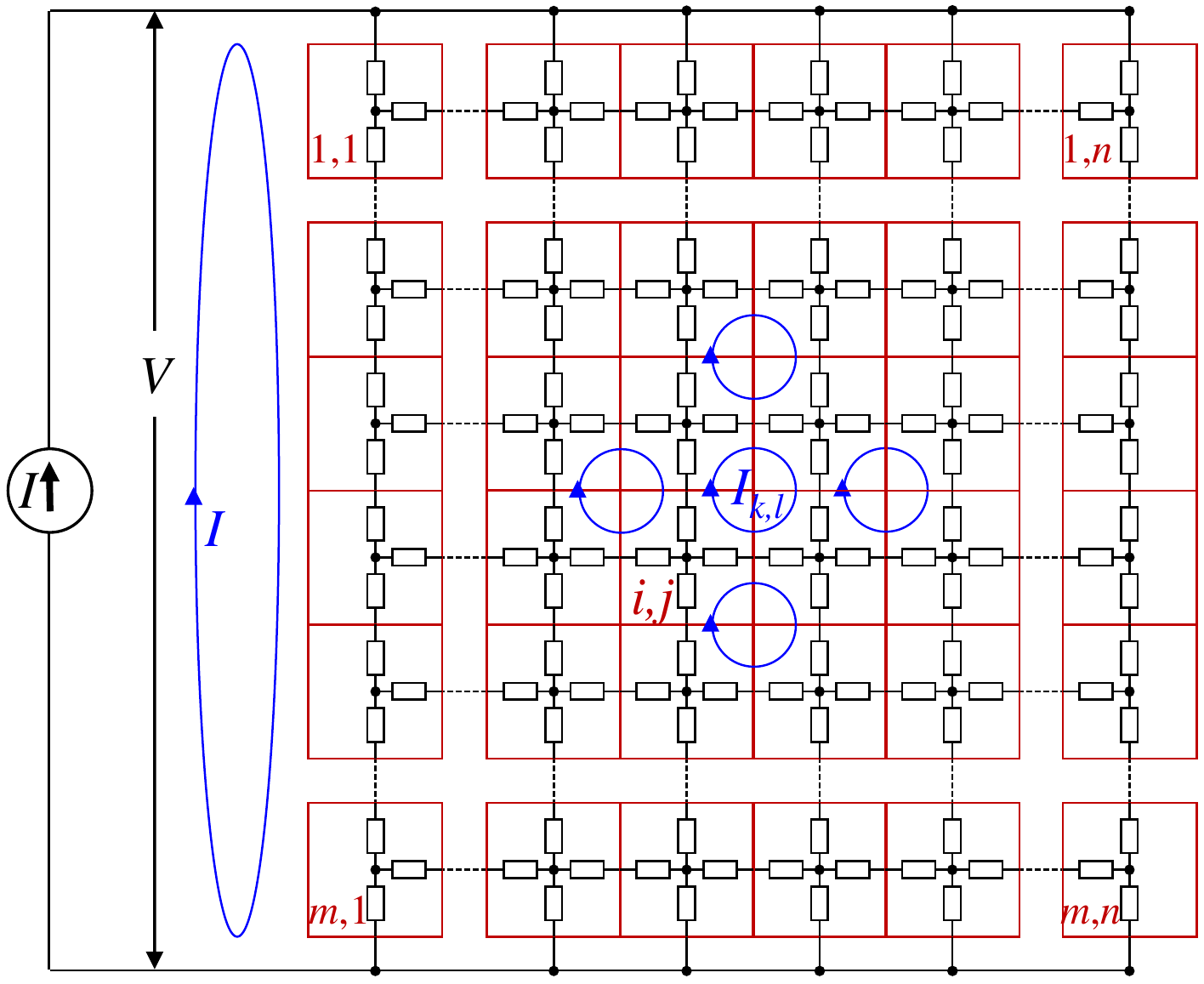}
\caption{Resistor network of $m\times n$ primitive cells used to approximate the resistance of the sample.
Each primitive cell (red squares) contains one node and four resistors (except for the columns $j=1$ and $j=n$ with only three resistors per primitive cell).
A mesh current $I_{k,l}$ (blue) flows in each of the essential meshes of the resistor network.
An additional mesh containing the current source is connected to the left side of the resistor network.  
\label{Fig_S_Resistor_Network}}
\end{figure}

To calculate the current distribution the sample is approximated by a 2-dimensional resistor network that is constructed from a square primitive cell with edge length $\Delta x=\Delta y$ that contains one node and four resistors.
The resulting resistor network, obtained by assembling $n$ primitive cells in $x$-direction and $m$ primitive cells in $y$-direction, is shown in Fig.~\ref{Fig_S_Resistor_Network}.
The four resistors within each primitive cell with index $i,j$ with $i\in \left\{1,m\right\}$ and $j\in\left\{1,n\right\}$ are assigned the same resistance $R_{i,j}$.
The resistors at the left and right edge of the resistor network are truncated, making the boundary insulating, and a perfectly conducting wire connects the bottom and top edge to the current source.
The loops formed by four neighbouring nodes and the resistors between them are called essential meshes.

\begin{figure}[b!]
\includegraphics[width=0.6\columnwidth]{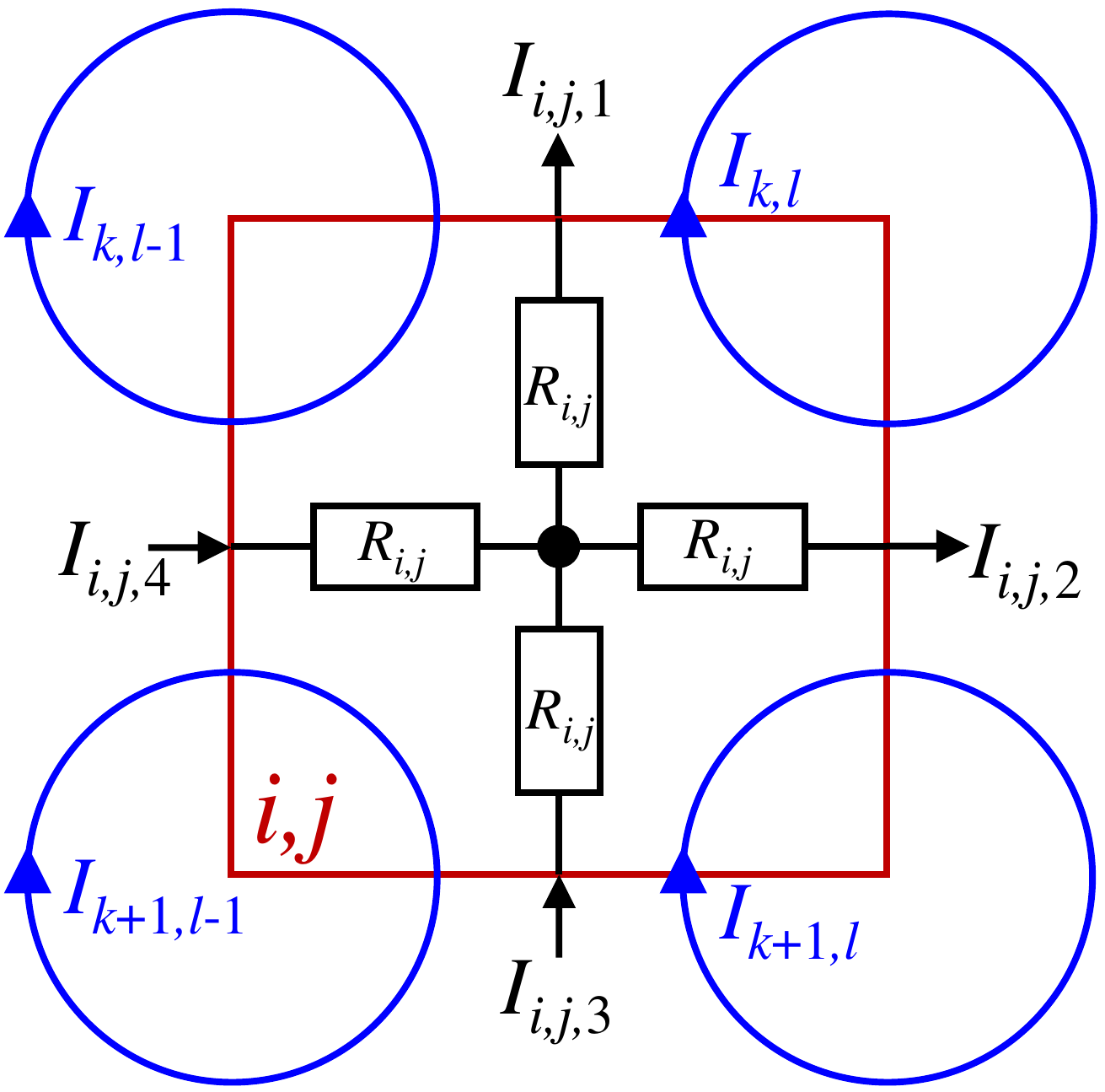}
\caption{
On the definition of the currents $I_{i,j,1}$ to $I_{i,j,4}$ flowing through the resistors in the primitive cell $i, j$, which are calculated from the adjacent mesh currents.
\label{Fig_S_ij-cell}}
\end{figure}

For the complete resistor network one obtains a set of $a=(n-1)(m+1)$ equations for $a$ essential meshes plus the equation for the current source mesh, which serves as a boundary condition.
This set of equations is assembled into a $\left(a+1\right)\times\left(a+1\right)$ matrix $\mathbf{EC}$ so that the equation system can be written in matrix form as

%
\begin{equation} 
\label{currentmatrix}
\mathbf{EC}\cdot
\begin{pmatrix}
I_{11}\\
\vdots\\
I_{1,n-1}\\
\vdots\\
I_{m+1,n-1}\\
I
\end{pmatrix}
=
\begin{pmatrix}
0\\
\vdots\\
0\\
\vdots\\
0\\
V
\end{pmatrix}~.
\end{equation}
%
Solving this equation system returns for a given bias current $I$ the voltage V across the network and the mesh currents, from which the current through each of the four resistors in the primitive cell $i,j$  can be calculated according to
%
\begin{align}
I_{i,j,1}&=I_{k,l}-I_{k,l-1}\nonumber\\
I_{i,j,2}&=I_{k+1,l}-I_{k,l}\nonumber\\
I_{i,j,3}&=I_{k+1,l}-I_{k+1,l-1}\nonumber\\
I_{i,j,4}&=I_{k+1,l-1}-I_{k,l-1}~,
\end{align}
%
where, by convention, currents flowing in positive $x$- and $y$-direction have positive sign.
These describe the current flow between adjacent primitive cells and are defined on the boundary between them (cf.~Fig.~\ref{Fig_S_ij-cell}).

Although it is not properly defined, it is desirable to evaluate the current density at the nodes.
This is calculated by interpolating the current densities at the boundaries of the primitive cell, which delivers the current densities $J_{i,j}^x$ in $x$-direction and $J_{i,j}^y$ in $y$-direction, as well as the magnitude of the current density $J_{i,j}^{\mathrm{norm}}$
%
\begin{align}
J_{i,j}^x&=\frac{I_{i,j,2}+I_{i,j,4}}{2\,\Delta y\, d_{\mathrm{f}}}\nonumber\\
J_{i,j}^y&=\frac{I_{i,j,1}+I_{i,j,3}}{2\,\Delta x\, d_{\mathrm{f}}}\nonumber\\
J_{i,j}^{\mathrm{norm}}&=\sqrt{\left(J_{i,j}^x\right)^2+\left(J_{i,j}^y\right)^2}~,
\end{align}
%
with the film thickness $d_{\mathrm{f}}$.
Note that these quantities approach the correct value only for $\Delta x,\,\Delta y\to 0$ and might deviate from the correct solution for finite dimensions of the primitive cell.
Therefore, they are only used for displaying the results and not as input for further calculations.
The power density $p_{i,j}$ generated through Joule heating in the primitive cell of volume $V_{i,j}$ is a quantity that will be used for modeling the thermal characteristics of the sample and is given by
%
\begin{equation}
p_{i,j}=\frac{P_{i,j}}{V_{i,j}}=\frac{R_{ij}}{\Delta x\, \Delta y\,d_{\mathrm{f}}}\left(I_{i,j,1}^2+I_{i,j,2}^2+I_{i,j,3}^2+I_{i,j,4}^2\right)~.
\end{equation}
%
For the reasons mentioned above, the currents through the boundaries are used to calculate the power dissipated in the resistors within the primitive cell.

\subsection{Definition of temperature-dependent resistivity}

To approximate the temperature dependence of the resistivity of the V$_2$O$_3$ film, the following assumptions are made:
(i) The resistivity $\rho_{\mathrm{met}}$ in the metallic phase is constant with respect to temperature and lateral position.
(ii) The resistivity $\rho_{\mathrm{ins}}$ deep in the insulating phase (at temperatures far below the MIT/IMT temperatures) is homogeneous and has the same temperature dependence as the globally measured $R(T)$.
(iii) For cool-down of the sample, the increase in resistivity from $\rho_{\mathrm{met}}$ to $\rho_{\mathrm{ins}}$ in the transition region is described by a function $f\left(T_{\mathrm{MIT}}\left(x,y\right)-T\right)$ that increases from 0 above the MIT temperature $T_{\mathrm{MIT}}$ to 1 at low temperatures.
Accordingly, the local resistivity for cool-down of the sample is described by the function
%
\begin{equation}
\rho_{\rm MIT}\left(x,y,T\right)=\rho_{\mathrm{ins}}\left(T\right)\,f\left(T_{\mathrm{MIT}}\left(x,y\right)-T\right)+\rho_{\mathrm{met}}~.
\end{equation}
%
(iv) Similarly, for warm-up of the sample, the decrease in resistivity in the transition region is described by the same function $f\left(T_{\mathrm{IMT}}\left(x,y\right)-T\right)$, that now has the difference between the IMT temperature $T_{\mathrm{IMT}}$ and the temperature as argument
%
\begin{equation}
\rho_{\rm IMT}\left(x,y,T\right)=\rho_{\mathrm{ins}}\left(T\right)\,f\left(T_{\mathrm{IMT}}\left(x,y\right)-T\right)+\rho_{\mathrm{met}}~.
\end{equation}
%
Since the transition temperatures vary spatially, this leads to an inhomogeneous distribution of the resistivity in the transition region, while the resistivity is homogeneous outside the transition region.
Note that the transition temperatures $T_{\mathrm{MIT}}\left(x,y\right)$ and $T_{\mathrm{IMT}}\left(x,y\right)$ that are used here to define the temperature dependence of the resistivity have been determined experimentally (cf.~Fig.~2), which should allow for a good  representation of the spatial dependence of the resistivity.

\begin{figure}[t]
\includegraphics[width=\columnwidth]{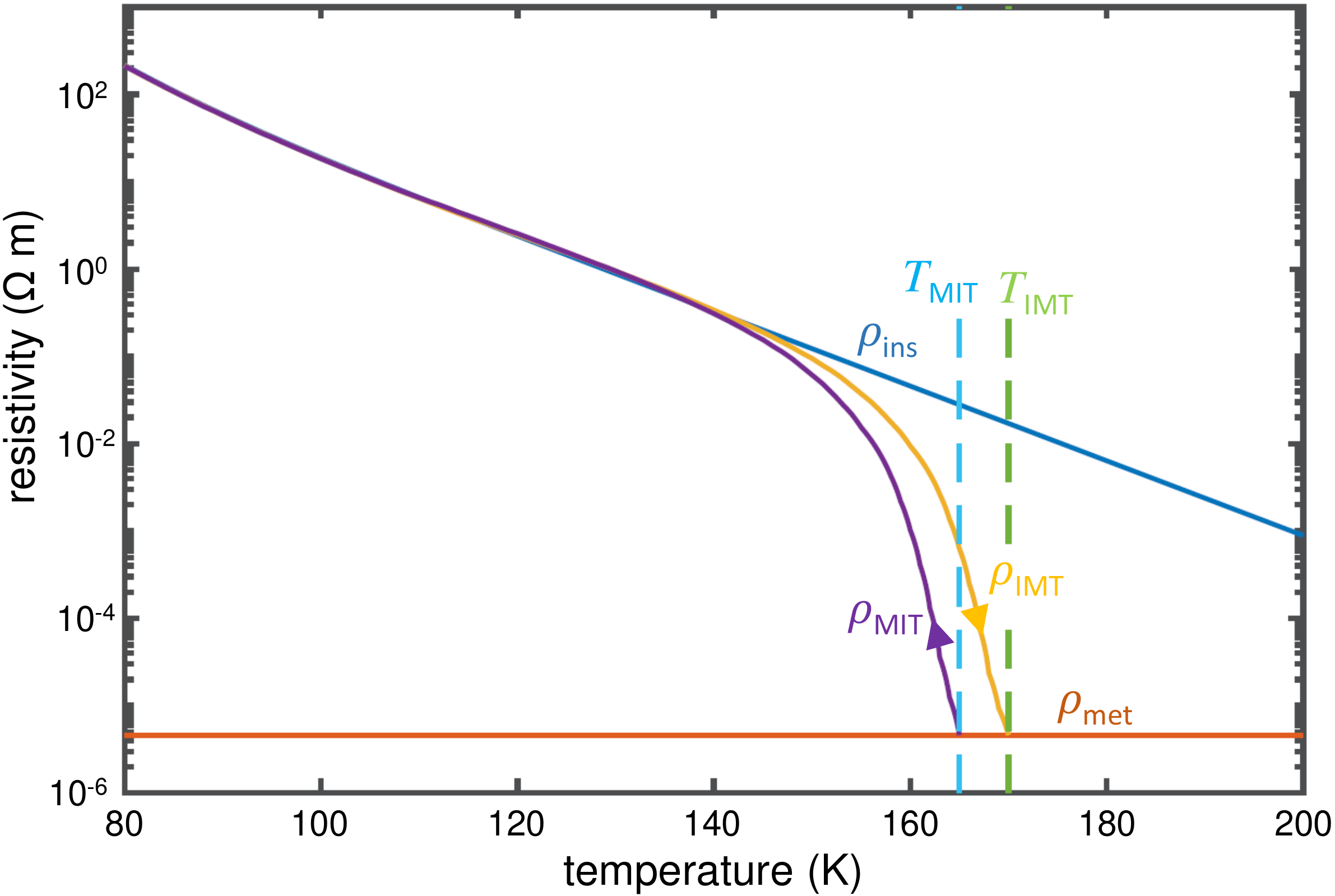}
\caption{
Modeling of the temperature dependence of the V$_2$O$_3$ resistivity.
For the cooling branch $\rho_{\rm MIT}$ increases from $\rho_\mathrm{met}$ to $\rho_\mathrm{ins}$ for $T<T_\mathrm{MIT}$ (blue dashed line).
In the heating branch $\rho_{\rm IMT}$ decreases towards $T_\mathrm{IMT}$ (green dashed line).
\label{fig:resVsT}}
\end{figure}

The resistivity in the metallic phase has been dereived from experimental data as $\rho_{\mathrm{met}}= 4.618\times 10^{-6}\,\mathrm{\Omega m}$.
The resistivity at low temperatures was obtained by fitting the measured $R(T)$ curve in the temperature range from 80\,K to 130\,K with the function 
%
\begin{equation}
\rho_{\mathrm{ins}}\left(T\right)=a\mathrm{e}^{bT}+c\mathrm{e}^{dT}~,
\end{equation}
%
with $a=4.66\times 10^8\,\Omega$m, $b=-0.1926\,$/K, $c=3.19\times 10^5\,\Omega$m, and $d=-0.09849\,$/K.
The function $f$ that describes the transition between the metallic and insulating phase is obtained by running the mesh current analysis with the matrix $R_{i,j}\left(T\right)$ as input for temperatures throughout the transition region and adjusting $f$ until the measured resistance is reproduced.
The resulting temperature-dependent resistivity is shown in Fig.~\ref{fig:resVsT}.

\subsection{Finite-difference approximation to heat equation}

Heat transfer in solids is described by the parabolic partial differential equation, known as the heat equation,
%
\begin{equation}
\rho \, c_P\frac{\partial T}{\partial t}-\bm\nabla \left(k\bm\nabla T\right)=\dot{q}_V~,
\end{equation}
%
with the density $\rho$, the heat capacity at constant pressure $c_P$, the thermal conductivity $k$, and the volumetric heat source $\dot{q}_V$.
In two dimensions and for a film thickness $d_{\mathrm{f}}$ this can be written as
%
\begin{equation}
\rho \,c_P\frac{\partial T}{\partial t}-\frac{\partial}{\partial x}\left(k_x\frac{\partial T}{\partial x}\right)-\frac{\partial}{\partial y}\left(k_y\frac{\partial T}{\partial y}\right)=\dot{q}_V~,
\end{equation} 
%
or, if the thermal conductivity $k=k_x=k_y$ is isotropic, as
%
\begin{equation}
\label{2Dheat}
\frac{\partial T}{\partial t}-\kappa\left(\frac{\partial^2 T}{\partial x^2}+\frac{\partial^2 T}{\partial y^2}\right)=\frac{\dot{q}_V}{\rho c_P}~,
\end{equation} 
%
with the thermal diffusivity $\kappa=k/\left(\rho c_P\right)$.
The partial derivatives in the heat equation can be approximated by finite differences.
The same discretization in space as for the mesh current analysis will be used and the variables are evaluated on the central node of the primitive cell with index $i,j$.
The time is discretized into time steps $\Delta t$ and an index $\tau$ is introduced describing the point in time.
The heat equation can be approximated in space and time using a number of finite-difference methods.
Here, the backward Euler method is used which, in contrast to the forward Euler or Crank-Nicolson method, is unconditionally stable and immune to oscillations.
For a node spacing of $\Delta x$ and $\Delta y$ and a time step $\Delta t$ the implicit discrete form of Eq.~\eqref{2Dheat} is given by
%
\begin{align}\label{2Dheatfin}
\frac{T_{i,j}^{\tau+1}-T_{i,j}^{\tau}}{\Delta t}
&-\kappa\Bigg(\frac{T_{i,j+1}^{\tau+1}-2T_{i,j}^{\tau+1}+T_{i,j-1}^{\tau+1}}{\left(\Delta x\right)^2}\\ \nonumber
&+\frac{T_{i+1,j}^{\tau+1}-2T_{i,j}^{\tau+1}+T_{i-1,j}^{\tau+1}}{\left(\Delta y\right)^2}\Bigg)
=\frac{\dot{q}_{i,j}^\tau}{\rho \,c_P}~.
\end{align}

At this point, an additional thermal coupling to the bath with temperature $T_{\mathrm{b}}$ is introduced which is a consequence of the heat flow to the coldfinger through the substrate of thickness $d_{\mathrm{s}}$ and thermal conductivity $k_{\mathrm{s}}$.
This can be modeled as a heat source $\dot{Q}_{\mathrm{b}}=k_{\mathrm{s}}\left(T_{\mathrm{b}}-T_{i,j}^{\tau+1}\right)\,\Delta x\Delta y/d_{\mathrm{s}}$ that is given by the power that is transfered to the bath through a cuboid with cross section $\Delta x\, \Delta y$ and length $d_{\mathrm{s}}$.
Accordingly, the volumetric heat source in Eq.~\eqref{2Dheatfin} is represented by the sum of Joule heating $p_{i,j}$ and a contribution of $\dot{Q}_{\mathrm{b}}$ lumped to the volume $V_{i,j}$
%
\begin{equation}
\dot{q}_{i,j}^\tau=p_{i,j}^\tau+\frac{\dot{Q}_{\mathrm{b}}}{V_{i,j}}=p_{i,j}^\tau+\frac{k_{\mathrm{s}}}{d_{\mathrm{f}}d_{\mathrm{s}}}\left(T_{\mathrm{b}}-T_{i,j}^{\tau+1}\right)~.
\end{equation}

It is further assumed, that the node spacing in $x$- and $y$-direction is equal ($\Delta x= \Delta y$).
With the thermal diffusivity to the bath $\kappa_{\mathrm{s}}=k_{\mathrm{s}}/\left(\rho_{\rm s}\,c_{P,{\rm s}}\right)$, Eq.~\eqref{2Dheatfin} becomes

\begin{widetext}
%
\begin{equation}
T_{i,j}^{\tau+1}-T_{i,j}^{\tau}-\frac{\kappa\,\Delta t}{\left(\Delta x\right)^2}\left(T_{i-1,j}^{\tau+1}+T_{i,j-1}^{\tau+1}-4T_{i,j}^{\tau+1}+T_{i,j+1}^{\tau+1}+T_{i+1,j}^{\tau+1}\right) 
=\frac{p_{i,j}^\tau\,\Delta t}{\rho \,c_P}+\frac{\kappa_{\mathrm{s}}\,\Delta t}{d_{\mathrm{f}}\,d_{\mathrm{s}}}\left(T_{\mathrm{b}}-T_{i,j}^{\tau+1}\right)~.
\end{equation}
%
Rearranging this equation so that terms at time $\tau+1$ are on the left hand side and terms at time $\tau$ are on the right hand side, and introducing the coefficients $A_{\mathrm{f}}=\kappa\,\Delta t/\left(\Delta x\right)^2$, $A_{\mathrm{b}}=\kappa_{\mathrm{s}}\,\Delta t/\left(d_{\mathrm{f}}\,d_{\mathrm{s}}\right)$, and $A_{\mathrm{q}}=\Delta t/\left(\rho\,c_P\right)$, results in
%
\begin{equation}
-A_{\mathrm{f}}\,T_{i-1,j}^{\tau+1}-A_{\mathrm{f}}\,T_{i,j-1}^{\tau+1}+\left(1+4A_{\mathrm{f}}+A_{\mathrm{b}}\right)T_{i,j}^{\tau+1}-A_{\mathrm{f}}\,T_{i,j+1}^{\tau+1} 
-A_{\mathrm{f}}\,T_{i+1,j}^{\tau+1}-A_{\mathrm{b}}T_{\mathrm{b}}
=T_{i,j}^{\tau}+A_{\mathrm{q}}\,p_{i,j}^\tau~.
\end{equation}
%
\end{widetext}
%
These equations are assembled into a $\left(m\cdot n+1\right)\times\left(m\cdot n+1\right)$ heat transfer matrix $\mathbf{HT}$ so that the equation system is given by 
%
\begin{equation}\label{heatmatrix}
\mathbf{HT}\cdot
\begin{pmatrix}
T_{11}^{\tau+1}\\
\vdots\\
T_{1n}^{\tau+1}\\
\vdots\\
T_{mn}^{\tau+1}\\
T_{\mathrm{b}}
\end{pmatrix}
=
\begin{pmatrix}
T_{11}^{\tau}+A_{\mathrm{q}}\,p_{1,1}^\tau\\
\vdots\\
T_{1n}^{\tau}+A_{\mathrm{q}}\,p_{1,n}^\tau\\
\vdots\\
T_{mn}^{\tau}+A_{\mathrm{q}}\,p_{m,n}^\tau\\
T_{\mathrm{bath}}
\end{pmatrix}~,
\end{equation}
%
where the last row contains the equation $T_{\mathrm{b}}= T_{\mathrm{bath}}$ which is used to set the bath temperature as boundary condition.

\begin{figure}[b]
\includegraphics[width=\columnwidth]{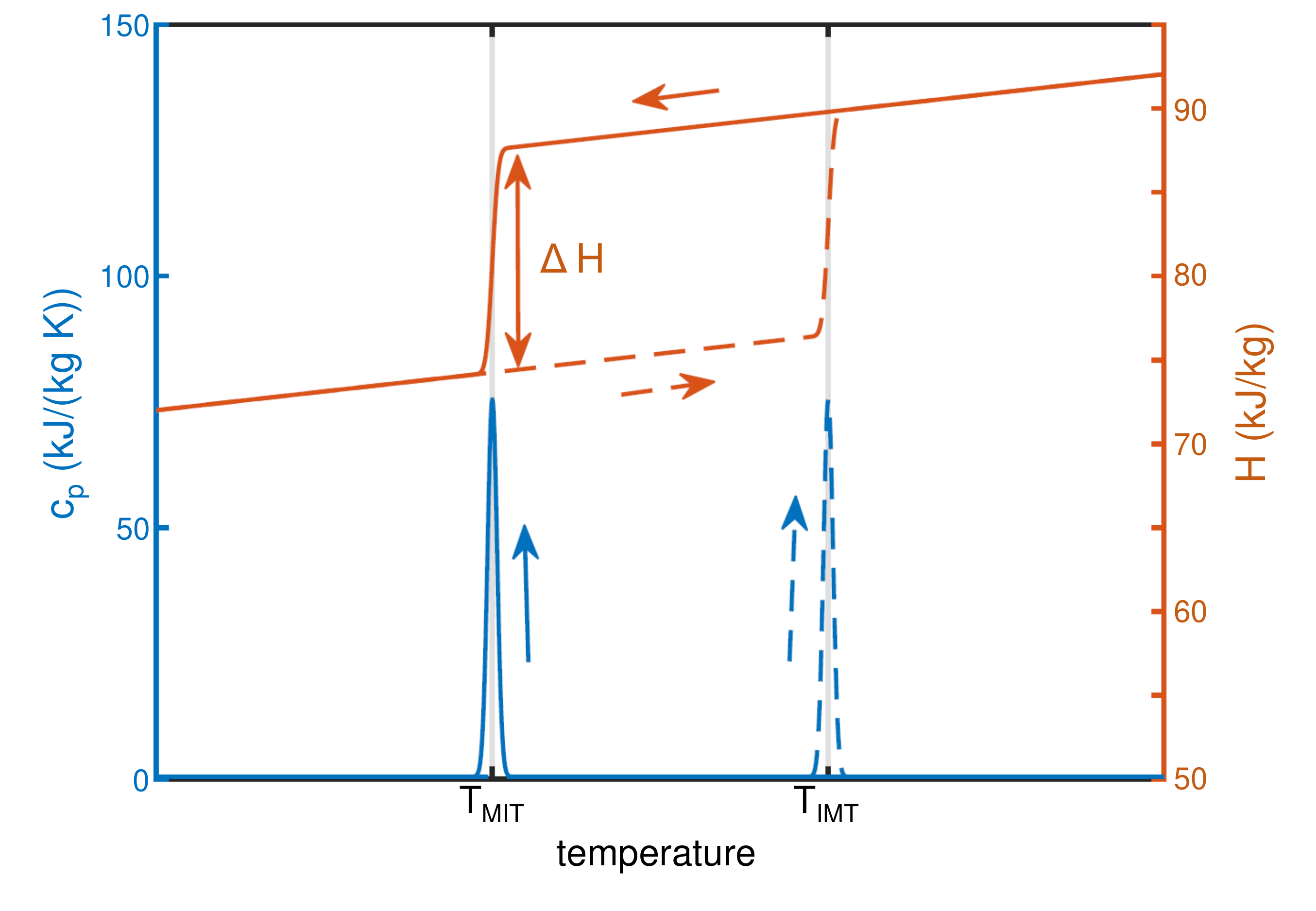}
\caption{
Temperature dependent heat capacity $c_P$ (blue curve) and enthalpy $H$ (red curve). The latent heat $\Delta H$ at the phase transition leads to a jump in enthalpy and a peak in heat capacity.
Broken lines show the IMT at $T_{\mathrm{IMT}}$, solid lines the MIT at $T_{\mathrm{MIT}}$. 
\label{fig:heatcap}}
\end{figure}

The metal-insulator phase transition in V$_2$O$_3$ is connected to the occurrence of a latent heat of approximately $\Delta H=2\,\mathrm{kJ/mol}\approx 13.3\,\mathrm{kJ/kg}$ that has to be supplied to the system to change from one phase to the other~\cite{Keer76}.
The latent heat, which is represented by a jump in enthalpy $H$, leads to a diverging heat capacity $c_P=\mathrm{d}H/\mathrm{d}T$ at the phase transition~\cite{Lyakh12}.
The latent heat is represented in the numerical model by introducing a temperature dependent heat capacity
%
\begin{equation}\label{heatcap}
 c_P\left(T\right)=c_{P0}+\frac{\Delta H}{\sqrt{\pi}\Delta T} e^{\frac{\left(T-T_{\mathrm{c}}\right)^2}{\Delta T^2}}~,
 \end{equation}
%
where a constant heat capacity $c_{P0}=0.45\,\mathrm{kJ/\left(kg\,K\right)}$ and a finite width $\Delta T=0.1\,\mathrm{K}$ of the phase transition around the transition temperature $T_{\mathrm{c}}$ is assumed.
The heat capacity according to Eq.~\eqref{heatcap} is shown in Fig.~\ref{fig:heatcap}.

\subsection{Simulation of electric breakdown in V$_2$O$_3$}

\begin{figure}[b]
\includegraphics[width=\columnwidth]{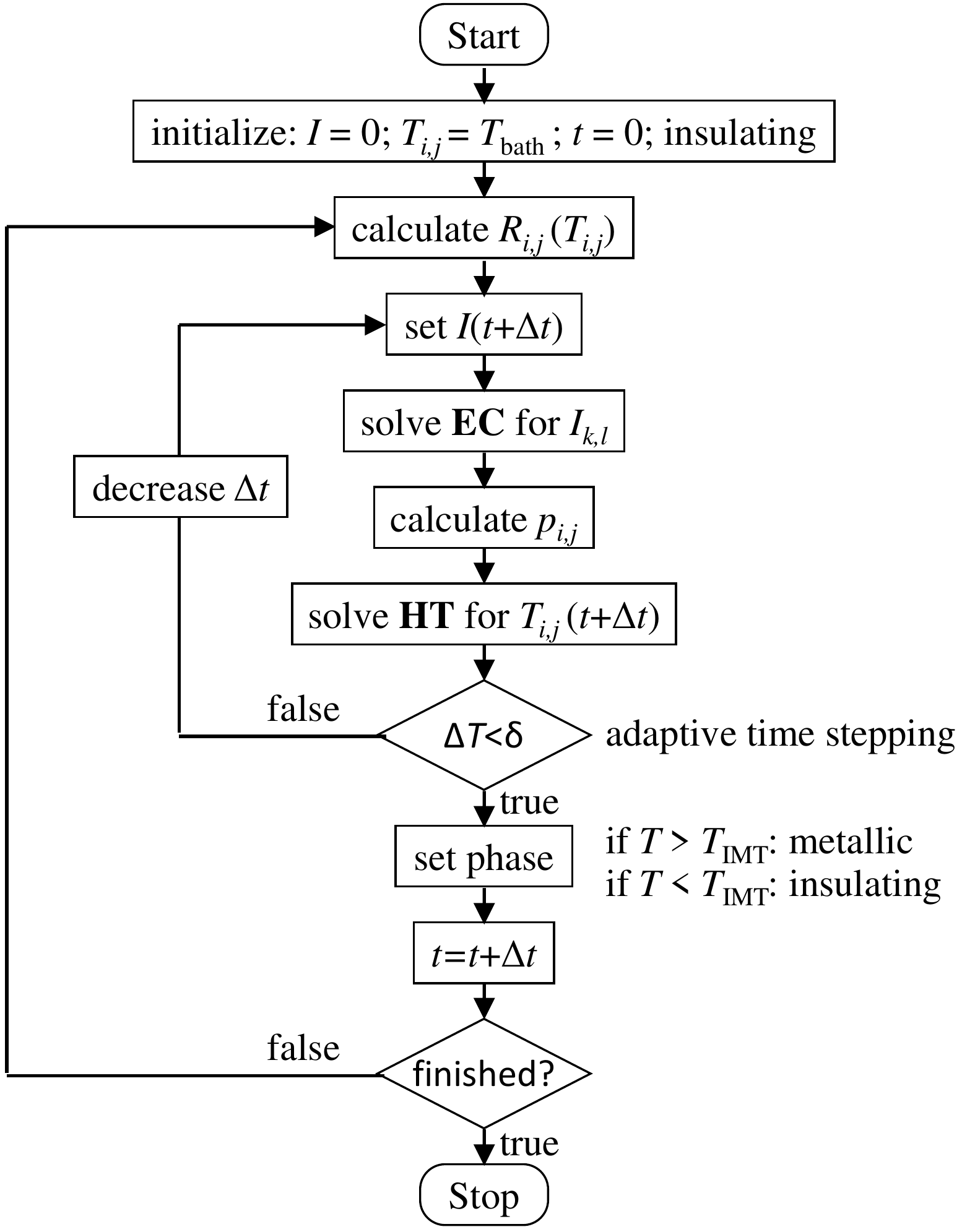}
\caption{
Flow chart for the simulation.
\label{fig:simuflow}}
\end{figure}

To simulate the electric breakdown in V$_2$O$_3$, the equations for the current distribution and heat transfer [Eq.~\eqref{currentmatrix} and Eq.~\eqref{heatmatrix}, respectively] are implemented and solved in MATLAB.
This is done by initializing the system at zero bias current ($I_{\mathrm{bias}}=0$) and at the bath temperature ($T_{i,j}=T_{\mathrm{bath}}$) for $t=0$ and slowly ramping the current to $I_{\mathrm{bias}}^{\mathrm{max}}=20\,\mathrm{mA}$ and back to $I_{\mathrm{bias}}^{\mathrm{end}}=0\,\mathrm{mA}$ over time with a rate that is much slower than the thermal dynamics of the sample, so that the system is modeled in quasi-static approximation.

Figure~\ref{fig:simuflow} shows the programs flow chart.
First, the system is initialized at the bath temperature, zero bias current and completely in the insulating phase.
Then, the $R_{i,j}$ for the initial temperature and phase are calculated.
The bias current is increased to the value at the next time step $t+\Delta t$.
The mesh current analysis is solved and the current distribution is calculated.
The Joule heating is calculated from $R_{i,j}$ and the current distribution and input into the heat equation, which is solved for the temperatures $T_{i,j}\left(t+\Delta t\right)$. 
Note that, due to the strong temperature dependence of resistivity and heat capacity, the system is highly nonlinear as it approaches the MIT or IMT.
To accurately capture the nonlinearities a sufficiently small time step $\Delta t$ needs to be used.
This is accomplished by using an adaptive time step.
If the maximum change in temperature 
%
\begin{equation}
\Delta T=\underset{\forall i,j}{\mathrm{max}} \left(| T_{i,j}\left(t+\Delta t\right)-T_{i,j}\left(t\right)|\right)
\end{equation}
%
is larger than the convergence criterion $\delta$, the time step is reduced and the temperature distribution is recalculated using the smaller time step.
If $\Delta T$ is far below the convergence criterion, the time step is increased.
When the convergence criterion is satisfied, the phase transition is evaluated:
the material is set to the metallic phase, where $T_{\mathrm{IMT}}$ is exceeded and to the insulating phase, where the temperature has fallen below $T_{\mathrm{MIT}}$.
Subsequently, the time is incremented and the calculated temperature and phase distribution are used as inputs to calculate the new $R_{i,j}$ values.
The program runs in a loop, progressing through time, until the bias current has reached its final value.

\bibliography{supplV2O3microscopy}